%% file: arxiv.tex
\pdfoutput=1
\documentclass[acmsmall,screen,nonacm]{acmart}

\usepackage{multirow}
\usepackage{tabularx}
\usepackage{makecell}
\usepackage{fontawesome5}
\newcommand{\staricon}{\raisebox{0.1ex}{\scalebox{0.78}{\faStar}}}
\newcommand{\langicon}[1]{\raisebox{-0.3ex}{\includegraphics[height=1.75ex]{figures/icons/#1}}}
\newcommand{\grouplabel}[1]{\rotatebox[origin=c]{90}{\shortstack{#1}}}

\usepackage{listings}
\usepackage{wrapfig}
\usepackage{tikz}
\usetikzlibrary{arrows.meta,positioning,calc}
\definecolor{lstkw}{HTML}{0033B3}
\definecolor{lststr}{HTML}{067D17}
\definecolor{lstcom}{HTML}{8C8C8C}
\definecolor{lstnum}{HTML}{9AA0A6}
\definecolor{lstbg}{HTML}{F7F8FA}
\definecolor{lstrule}{HTML}{D0D7DE}
\definecolor{lstmark}{HTML}{FFE9A8}
\definecolor{lstaccent}{HTML}{C62828}
\lstdefinestyle{paper}{
  language=Python, basicstyle=\ttfamily\scriptsize, columns=fullflexible, keepspaces=true,
  keywordstyle=\color{lstkw}\bfseries, stringstyle=\color{lststr}, commentstyle=\color{lstcom}\itshape,
  morekeywords={self}, emph={startswith,replace}, emphstyle=\color{lstkw},
  numbers=left, numberstyle=\tiny\color{lstnum}, numbersep=5pt, xleftmargin=2.1em, framexleftmargin=1.9em,
  breaklines=true, breakatwhitespace=true, postbreak=\mbox{\ \ }, backgroundcolor=\color{lstbg},
  rulecolor=\color{lstrule}, framerule=0.5pt, aboveskip=0pt, belowskip=0pt,
  escapeinside={(*@}{@*)}, showstringspaces=false, upquote=true}
\DeclareRobustCommand{\callout}[1]{\tikz[baseline=(c.base)]\node[circle,fill=lstaccent,text=white,inner sep=0.7pt,minimum size=1.35ex,font=\sffamily\bfseries\fontsize{5.5}{5.5}\selectfont](c){#1};}
\DeclareRobustCommand{\lstmark}[1]{{\setlength{\fboxsep}{0.4pt}\colorbox{lstmark}{\vphantom{Ay}#1}}}

\usepackage{etoolbox}
\usepackage{enumitem}
\setlist{itemsep=1pt, topsep=3pt, parsep=0pt, partopsep=0pt}
\AtBeginDocument{\captionsetup{skip=2pt}}
\appto\normalsize{%
  \setlength{\abovedisplayskip}{4pt plus 2pt minus 1pt}%
  \setlength{\belowdisplayskip}{4pt plus 2pt minus 1pt}%
  \setlength{\abovedisplayshortskip}{2pt plus 1pt}%
  \setlength{\belowdisplayshortskip}{2pt plus 1pt}}

\newcommand{\finding}[2]{%
  \par\vspace{1pt}\noindent\fbox{\parbox{\dimexpr\linewidth-2\fboxsep-2\fboxrule\relax}{%
    \textbf{Finding #1:} #2}}\par\vspace{1pt}}

\newcommand{\corrmark}{\raisebox{-0.1ex}{\scalebox{0.82}{\faEnvelope[regular]}}}
\makeatletter
\def\@mkauthors{\begingroup
  \hsize=\textwidth
  \global\setbox\mktitle@bx=\vbox{\noindent\unvbox\mktitle@bx\par\medskip
    \noindent{\raggedright\large\sffamily
      \mbox{Shihao Weng\textsuperscript{1,$*$}},
      \mbox{Longtian Wang\textsuperscript{2,$*$}},
      \mbox{Yang Feng\textsuperscript{1,\corrmark}},
      \mbox{Mingfei Cheng\textsuperscript{3}},
      \mbox{Jia Liu\textsuperscript{1}},
      \mbox{Chao Shen\textsuperscript{2}},
      \mbox{He Ye\textsuperscript{4}},
      \mbox{Federica Sarro\textsuperscript{4}},
      \mbox{Xiaofei Xie\textsuperscript{3}}\par}\smallskip
    \noindent{\raggedright\small
      \mbox{\textsuperscript{1}Nanjing University, Nanjing, China}\quad
      \mbox{\textsuperscript{2}Xi'an Jiaotong University, Xi'an, China}\quad
      \mbox{\textsuperscript{3}Singapore Management University, Singapore}\quad
      \mbox{\textsuperscript{4}University College London, London, United Kingdom}\par}\medskip}%
  \endgroup}
\makeatother

\begin{document}

\title[When Code Becomes an LLM Call]{When Code Becomes an LLM Call: An Empirical Study of Call-Site Contracts in Agentic Systems}

\author{Shihao Weng}
\affiliation{\institution{Nanjing University}\city{Nanjing}\country{China}}
\email{shweng@smail.nju.edu.cn}
\author{Longtian Wang}
\affiliation{\institution{Xi'an Jiaotong University}\city{Xi'an}\country{China}}
\email{w976625943@stu.xjtu.edu.cn}
\author{Yang Feng}
\affiliation{\institution{Nanjing University}\city{Nanjing}\country{China}}
\email{fengyang@nju.edu.cn}
\author{Mingfei Cheng}
\affiliation{\institution{Singapore Management University}\country{Singapore}}
\email{snowbirds.mf@gmail.com}
\author{Jia Liu}
\affiliation{\institution{Nanjing University}\city{Nanjing}\country{China}}
\email{liujia@nju.edu.cn}
\author{Chao Shen}
\affiliation{\institution{Xi'an Jiaotong University}\city{Xi'an}\country{China}}
\email{chaoshen@mail.xjtu.edu.cn}
\author{He Ye}
\affiliation{\institution{University College London}\city{London}\country{United Kingdom}}
\email{he.ye@ucl.ac.uk}
\author{Federica Sarro}
\affiliation{\institution{University College London}\city{London}\country{United Kingdom}}
\email{f.sarro@ucl.ac.uk}
\author{Xiaofei Xie}
\affiliation{\institution{Singapore Management University}\country{Singapore}}
\email{xfxie@smu.edu.sg}
\renewcommand{\shortauthors}{Weng et al.}
\authorsaddresses{$^*$Both authors contributed equally to this research. \textsuperscript{\corrmark}Corresponding author.\newline
  Authors' Contact Information:
  Shihao Weng, \href{mailto:shweng@smail.nju.edu.cn}{shweng@smail.nju.edu.cn};
  Longtian Wang, \href{mailto:w976625943@stu.xjtu.edu.cn}{w976625943@stu.xjtu.edu.cn};
  Yang Feng, \href{mailto:fengyang@nju.edu.cn}{fengyang@nju.edu.cn};
  Mingfei Cheng, \href{mailto:snowbirds.mf@gmail.com}{snowbirds.mf@gmail.com};
  Jia Liu, \href{mailto:liujia@nju.edu.cn}{liujia@nju.edu.cn};
  Chao Shen, \href{mailto:chaoshen@mail.xjtu.edu.cn}{chaoshen@mail.xjtu.edu.cn};
  He Ye, \href{mailto:he.ye@ucl.ac.uk}{he.ye@ucl.ac.uk};
  Federica Sarro, \href{mailto:f.sarro@ucl.ac.uk}{f.sarro@ucl.ac.uk};
  Xiaofei Xie, \href{mailto:xfxie@smu.edu.sg}{xfxie@smu.edu.sg}.}

\begin{abstract}
Software quality has long rested on specifications that state what code must do, and on testing and verification that check that the code does it. Agentic systems change this foundation: they replace parts of their logic, such as choosing the next tool, judging whether an action is allowed, or writing code to run, with calls to a large language model (LLM). The program still relies on what this logic must guarantee, but the model's behavior is not fixed by code, varies across runs, and can be steered by its input. The guarantee must therefore come from the program around each \emph{LLM call site}, as a contract of pre-conditions, prompt constraints, and post-conditions. We present an empirical study of these contracts in 496 LLM call sites of 40 popular open-source agentic systems (13.4\,MLOC, four languages). Using a multi-agent pipeline validated by 17,428 human judgments, we trace where every model output flows, derive an expected contract for each call site without access to its code, and compare it with the contract that the code implements. We find that 48\% of LLM outputs reach control flow or execution, and 73\% of these never reach a person. Programs enforce 78\% of their constraints in code, but in the prompt, 67\% of the constraints are only requested from the model. Although 82\% of call sites are checked before and after the call, 63\% of the expected constraints are unmet and only 5.3\% of call sites satisfy their whole contract; ten recurring patterns explain 99\% of the gaps. 38 mechanisms behind these gaps trigger at runtime, and a small code fix that enforces the expected constraint stops all 38. Our results show that the presence of checks says little about their adequacy, and we derive implications for agent developers, framework maintainers, and researchers.
\end{abstract}



\maketitle

\vspace{-4pt}
\section{Introduction}
\label{sec:intro}

In conventional software, quality rests on a well-understood foundation: developers specify what each unit of code must do, as pre- and post-conditions, types, assertions, or tests, and use testing and verification to check that the code does it~\cite{floyd1967assigning,hoare1969axiomatic,meyer1992applying,leavens2006jml}. Callers then rely on this specification without checking it again. The foundation works because a unit's behavior is fixed by its code.

Agentic systems change this foundation. Coding agents such as Codex, Gemini CLI, and OpenHands~\cite{wang2025openhands} edit repositories and run commands on developers' machines, and personal assistants read messages and act on users' accounts. Each of them is an ordinary program, but parts of its logic are delegated to a large language model (LLM)~\cite{yao2023react,schick2023toolformer,wang2024survey}. Where a conventional program would call a parser, a model extracts the user's intent; where it would call a dispatcher, a model chooses the next tool; where it would call a policy check, a model judges whether an action is allowed. We call each such place an \emph{LLM call site}: a location in the code where the program builds a request, calls a model, and acts on the reply.

Delegating logic to a model does not remove the need for its specification. The program still relies on what the logic must guarantee: that the chosen tool was offered, that the verdict is allowed, that the generated command is safe to run. The model does not establish these guarantees. Its behavior is not fixed by code, varies across runs, and can be steered by its input, through a model error~\cite{ji2023hallucination} or an injected instruction~\cite{perez2022ignore,greshake2023indirect,liu2023promptinjection}, and testing can sample but not establish this behavior~\cite{barr2015oracle}. The specification therefore moves into the program around the call, which must enforce a contract $\{P\}\,C\,\{Q\}$ before the call, in the request, and after the reply. For an LLM call, the post-condition $Q$ never holds by construction; it holds only where the program checks it. Whether an agentic system is reliable and secure depends on whether each call site enforces the contract it needs.

\begin{wrapfigure}{r}{0.52\textwidth}
  \vspace{-\intextsep}
  \centering
  \input{figures/motivating-example}
  \caption{A file read by Codewhale reaches a real process through three weak checks.}
  \label{fig:motivating}
  \vspace{-\intextsep}
\end{wrapfigure}
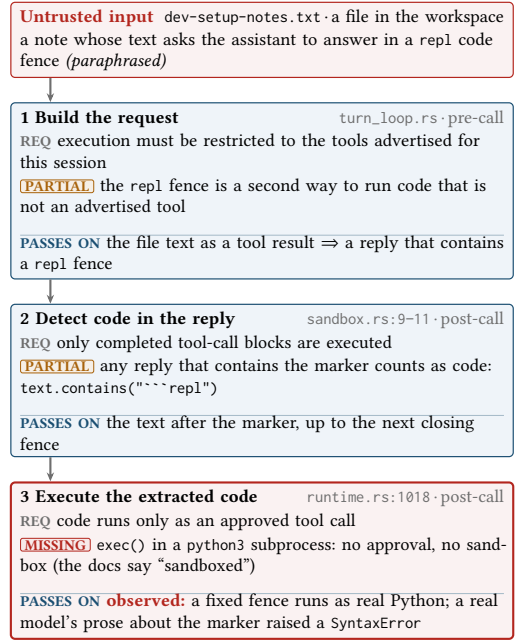

Little is known about these contracts. Benchmarks evaluate models and agents end to end~\cite{jimenez2024swebench,liu2024agentbench,yao2024taubench,debenedetti2024agentdojo}: they show whether a task fails or an attack succeeds, but not which check was missing. Defenses harden the model or its input~\cite{chen2025struq,hines2024spotlighting,debenedetti2025camel}, and empirical studies of LLM-based software examine how developers write prompts~\cite{liang2025prompts,tafreshipour2025prompting} or why agent workflows fail~\cite{cemri2025mast}. None of them asks the questions that the shift from code to model calls raises: which logic do programs delegate to LLMs, how do they specify and enforce what this logic must do, and is that enough?

The last question is hard for a principled reason. Reading the code shows which checks exist, but a missing check leaves no trace in the code, and specification mining~\cite{ammons2002mining,ernst2007daikon} recovers what the code does and cannot tell what it should do. Measuring adequacy therefore needs a reference that is independent of the implementation. Figure~\ref{fig:motivating} shows why the presence of a check is not enough. Codewhale runs a model reply as Python whenever it contains a \texttt{\textasciigrave\textasciigrave\textasciigrave repl} fence, without approval or sandbox, although its documentation calls the feature ``sandboxed''. The program does check the reply, but only whether it contains code, never whether running it was approved.

This paper presents an empirical study of the contracts of LLM call sites. We analyze 496 call sites in 40 popular open-source agentic systems that span six task domains, four languages, and 13.4\,MLOC. Our study has three stages (Figure~\ref{fig:overview}). Stage~I locates the call sites, splits each reply into typed outputs, and traces their consumers. Stage~II reconstructs two contracts per call site: an \emph{expected contract}, derived from a code-free functional specification by an agent that never sees the code, and an \emph{implemented contract}, extracted from the code by an agent that never sees the expected contract. Comparing the two exposes missing and weak constraints that code reading alone cannot find. Both stages run as a multi-agent system, and three reviewers check every step with 17,428 judgments (Gwet's AC1 0.91 to 0.99). Stage~III tests whether the gaps trigger at runtime and whether enforcing the expected constraint prevents them. We answer four research questions:

\begin{itemize}
  \item \textbf{RQ1: Which logic do agentic systems delegate to LLM calls, and where do the outputs flow?} 48\% of LLM outputs reach control flow or execution, and 71\% are used only by the program or by other models.
  \item \textbf{RQ2: How do agentic systems specify and enforce the contracts of their LLM calls?} Programs enforce 78\% of their constraints in code, yet 67\% of prompt-phase constraints are only requested from the model.
  \item \textbf{RQ3: How adequate are the implemented contracts, and where do they fall short?} 63\% of the expected constraints are unmet, and although 82\% of call sites check before and after the call, only 5.3\% satisfy their whole contract; ten patterns explain 99\% of the gaps.
  \item \textbf{RQ4: What are the runtime consequences of contract gaps, and does enforcing the missing constraint prevent them?} For 114 gaps, the missing constraint lies on a threat chain; 38 of their 60 mechanisms trigger at runtime, and enforcing the expected constraint in code stops all 38 exploits.
\end{itemize}

\noindent In summary, this paper makes the following contributions:
\begin{itemize}
  \item The first large-scale empirical study, to our knowledge, of how agentic systems specify and enforce the logic they delegate to LLM calls.
  \item A contract view of LLM calls and a blind, multi-agent reconstruction method that measures missing constraints against a reference independent of the code.
  \item A catalog of ten recurring gap patterns and concrete implications for agent developers, framework maintainers, and researchers.
  \item A validated dataset of 496 call sites, 917 typed outputs, 5,017 implemented and 2,699 expected constraints, with source evidence and review records.
\end{itemize}

\vspace{-4pt}
\section{Background}
\label{sec:background}

\subsection{LLM Call Sites and Their Outputs}
\label{sec:bg-callsites}

An agentic system is a program that calls one or more LLMs while it runs and acts on their replies. An \emph{LLM call site} $s$ is a location in the program that sends a request to a model and receives the reply. One reply often carries several results with different uses. We therefore split the reply of $s$ into a set of \emph{typed outputs} $O(s)$. Each output $o$ has one \emph{operation} $\mathit{op}(o)$, which describes what the model is asked to do, and a set of \emph{consumption edges} $E(o)$, one for each kind of place where the program uses $o$. Table~\ref{tab:taxonomy} defines ten operations and seven edge types. We call an output \emph{decision-relevant} when some consumer lets it select or become behavior:
\begin{equation}
  \mathit{dec}(o) \iff E(o) \cap \{\texttt{control-flow}, \texttt{executed}\} \neq \emptyset .
  \label{eq:dec}
\end{equation}
A call site is decision-relevant if one of its outputs is. Decision relevance is a reachability property: the output can reach a decision, whether or not it changes a given run.

A call site may also serve several responsibilities, for example when one helper plans for one caller and summarizes for another. Because each responsibility needs its own contract, we analyze contracts per \emph{responsibility unit} $u = (s, r)$, a call site together with one responsibility $r$.

\vspace{-4pt}
\subsection{A Contract View of LLM Calls}
\label{sec:bg-contract}

Following Hoare logic~\cite{hoare1969axiomatic} and design by contract~\cite{meyer1992applying,leavens2006jml,barnett2005spec}, we view a responsibility unit as the command $C = M(\pi(x))$: the program renders its state $x$ into a request with a prompt template $\pi$, and the model $M$ answers. Its contract has three phases:
\begin{equation}
  K(u) = K_{\mathrm{pre}}(u) \cup K_{\mathrm{prompt}}(u) \cup K_{\mathrm{post}}(u).
  \label{eq:contract}
\end{equation}
Pre-condition constraints restrict when and with which inputs the call is made, for example that the target ticker is known. Prompt constraints shape the request, such as instructions, output schemas, and fixed sampling parameters. Post-condition constraints restrict which replies the program accepts before a consumer uses them, for example that a tool call names an offered tool.

Each constraint $c$ also has an \emph{assurance level} that states how strongly the program makes it hold:
\begin{equation}
  \alpha(c) \in \{\textsf{prompt-only} \prec \textsf{code-checked} \prec \textsf{capability-bound}\}.
  \label{eq:assurance}
\end{equation}
A prompt-only constraint is stated to the model and never verified. A code-checked constraint is tested by the program, which rejects or repairs a violating value. A capability-bound constraint cannot be violated because the unit lacks the capability, as when a tool is never exposed to the model. The order reflects who can break the constraint: a prompt-only constraint fails whenever the model errs or injected text tells it otherwise, a code-checked one fails only through a bug in the check, and a capability-bound one does not fail at all.

We distinguish the contract that a unit needs from the one it has. The \emph{expected contract} $K^{E}(u)$ contains the constraints that the unit's responsibility requires, each with a phase and a required assurance level. The \emph{implemented contract} $K^{I}(u)$ contains the constraints that the code enforces, each with its phase, mechanism, assurance level, and source evidence.

\vspace{-4pt}
\section{Study Design}
\label{sec:design}

\begin{figure*}[t]
  \centering
  \includegraphics[width=0.80\textwidth]{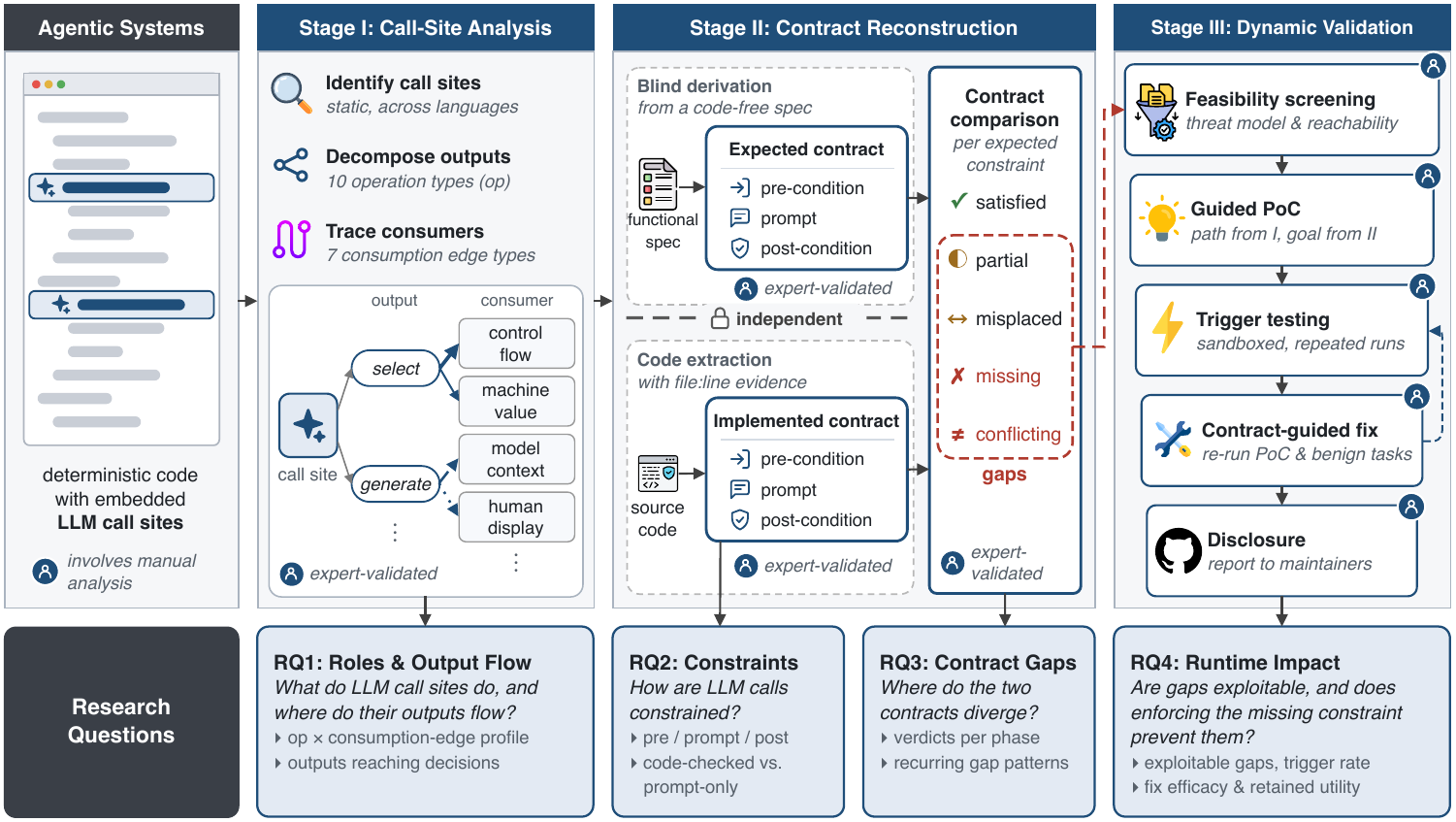}
  \caption{Overview of the study. The person icon marks steps with manual analysis or expert validation.}
  \label{fig:overview}
\end{figure*}

Figure~\ref{fig:overview} gives an overview. Stage~I (Section~\ref{sec:design-stage1}) answers RQ1, Stage~II (Section~\ref{sec:design-stage2}) answers RQ2 and RQ3, and Stage~III (Section~\ref{sec:design-stage3}) answers RQ4. LLM agents perform the analysis steps of Stages~I and~II, and human reviewers check every step before its output is used (Section~\ref{sec:design-reliability}). All agents of Stages~I and~II run on GPT-5.6 Sol, and Stage~III is performed manually.

\vspace{-4pt}
\subsection{Research Questions}
\label{sec:design-rqs}

The research questions follow the shift from code to model calls, from the delegated logic (RQ1) to its contracts (RQ2), their adequacy (RQ3), and their runtime consequences (RQ4). We keep RQ2 and RQ3 apart: RQ2 describes practice from code evidence alone, so its results do not depend on our notion of what a call site should check, and RQ3 judges it against the expected contract.

\vspace{-4pt}
\subsection{Subject Systems}
\label{sec:design-subjects}

\input{tables/subjects}

We study the 40 open-source agentic systems in Table~\ref{tab:subjects}, selected from popular GitHub agent repositories. A system qualifies if it is a concrete agent whose core logic is public, if an LLM takes part in its decisions at runtime, if the agent can call tools, act on an environment, change state, or choose what runs next, and if the inputs, outputs, and downstream uses of its LLM calls can be traced in the code. We exclude plain chatbots, question-answering and RAG systems that only produce text, agent frameworks and SDKs without a concrete agent, demos, and benchmark components such as user simulators, evaluators, and LLM judges. All systems have more than 5k GitHub stars (median 29k), and 30 date from 2024 or later. We group the systems by their documented primary task into six domains that follow the task families of agent surveys and benchmarks~\cite{wang2024survey}: software engineering~\cite{jimenez2024swebench}, computer and web use~\cite{xie2024osworld,zhou2024webarena}, research and science~\cite{chan2025mlebench}, general-purpose assistance~\cite{mialon2024gaia}, personal assistants that keep long-term memory and act on users' channels, and domain-specific agents for finance, security testing, data analysis, and embodied tasks.

The systems span 13.4\,MLOC of production code (non-blank source lines without tests, vendored, and generated code), each analyzed at a fixed commit. In Table~\ref{tab:subjects}, \emph{Lang} gives the language of the call sites, \emph{Cand.} counts the candidate sites we examined, and \emph{Sites} the confirmed call sites.

\vspace{-4pt}
\subsection{Stage I: Call-Site Analysis}
\label{sec:design-stage1}

Stage~I builds the output profile of every call site in three steps. Each step is performed by an LLM agent in an isolated context, and each claim that an agent makes must cite the source lines that support it. (1)~\emph{Identifying call sites.} Static front ends built on tree-sitter, combined with the parser of each language, locate candidate sites that can reach a model API or client library, following wrapper functions across files. Static analysis covers all 40 systems without running third-party code and reaches paths that tests rarely exercise. An identification agent then inspects each of the 1,226 candidates and decides whether it is a real model invocation, a wrapper, or a relay. Attributing wrappers and relays to the call sites they serve yields 496 call sites. (2)~\emph{Decomposing outputs.} A decomposition agent splits the reply of each call site into typed outputs and labels each with its operation, judged from the prompt and the API form, which yields 917 typed outputs. (3)~\emph{Tracing consumers.} A tracing agent follows each output through the program to the places that use it and labels each use with an edge type and its source location, which yields 1,222 consumption relations.

\input{tables/taxonomy}

\emph{Classification scheme.} Table~\ref{tab:taxonomy} lists the two dimensions. The operation depends only on what the model is asked to do, and the edge type depends only on what the program does with the result. The same \texttt{generate} output may thus be shown to a user or parsed into a tool call. Operations fall into five groups by their output space, from open-ended text to bounded choices, and edges fall into four consumer kinds: \emph{Control}, \emph{Data}, \emph{Model}, and \emph{Human}. Each category has a boundary rule against its nearest neighbor, such as whether content is copied from the input (\texttt{extract}) or created (\texttt{generate}). We developed the scheme on six pilot systems and froze it before measuring.

\vspace{-4pt}
\subsection{Stage II: Contract Reconstruction}
\label{sec:design-stage2}

\begin{figure*}[t]
  \centering
  \includegraphics[width=\textwidth]{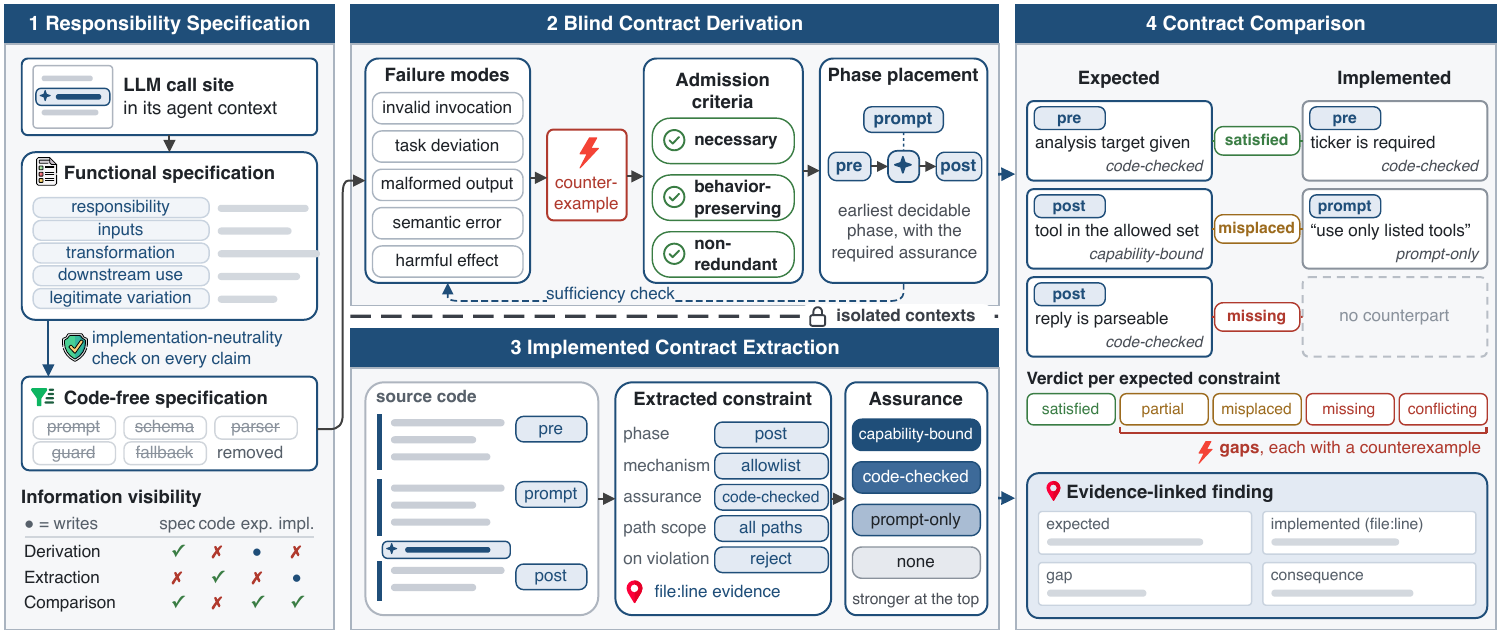}
  \caption{Multi-agent contract reconstruction for one responsibility unit.}
  \label{fig:contract}
\end{figure*}

Stage~II reconstructs the expected and the implemented contract of every responsibility unit and compares them. Figure~\ref{fig:contract} shows the multi-agent system~\cite{wu2024autogen,hong2024metagpt} that does so. Four agents run in separate, clean contexts, and each sees only the inputs allowed by the information-visibility matrix in the figure. With $\mathrm{ctx}(u)$ the call site in its agent context and $\mathrm{code}(u)$ its source code, we compute
\begin{equation}
\begin{aligned}
  \sigma(u) &= A_{\mathrm{spec}}\big(\mathrm{ctx}(u)\big), &
  K^{E}(u) &= A_{\mathrm{exp}}\big(\sigma(u)\big), \\
  K^{I}(u) &= A_{\mathrm{impl}}\big(\mathrm{code}(u)\big), &
  V(u) &= A_{\mathrm{cmp}}\big(\sigma(u), K^{E}(u), K^{I}(u)\big).
\end{aligned}
  \label{eq:pipeline}
\end{equation}
The derivation agent $A_{\mathrm{exp}}$ never sees the code, and the extraction agent $A_{\mathrm{impl}}$ never sees the expected contract. This separation is the core of our design: an analyst who reads the code first tends to inherit its omissions. Because $K^{E}$ depends on the code only through $\sigma$, it can reveal constraints the implementation lacks, and agreement between $K^{E}$ and $K^{I}$ cannot come from shared input. The specification is thus the only channel from the code to $K^{E}$, and we guard it twice: the specification agent tests it for mechanism independence, and code-blind reviewers recheck it (Section~\ref{sec:design-reliability}).

\emph{Specification agent.} $A_{\mathrm{spec}}$ reads the call site in its agent context, because a responsibility cannot be stated without knowing what the call site serves, and writes a functional specification: the responsibility, its inputs, its transformation, the downstream use of its outputs, and their legitimate variation. It reads prompts, parsers, and guards only as evidence of data flow and never copies their rules, fields, checks, or fallbacks. Before emitting $\sigma(u)$, it applies a mechanism-independence test: every statement must remain true if the current prompt format, schema, parser, guard, and fallback were replaced. The resulting $\sigma(u)$ states what the unit must achieve and not how the code does it.

\emph{Derivation agent.} $A_{\mathrm{exp}}$ receives only $\sigma(u)$. It enumerates how the unit can fail (invalid invocation, task deviation, malformed output, semantic error, harmful effect) and proposes constraints that prevent these failures. A constraint $c$ is admitted only if it is \emph{necessary}, shown by a concrete counterexample in which a reachable failure occurs without it; \emph{behavior-preserving}, rejecting no legitimate variation in $\sigma(u)$; and \emph{non-redundant}, not implied by the other admitted constraints. The agent places each constraint in the earliest phase that can decide it, assigns the assurance level that the failure requires, and finally adds constraints for any failure mode left uncovered. The criteria keep the expected contract minimal, so every gap points to a constraint that the unit needs.

\emph{Extraction agent.} $A_{\mathrm{impl}}$ receives only the source code of the unit and of the paths that reach it. It extracts every constraint enforced before, within, or after the call, with its phase, mechanism, assurance level, the paths it covers, its behavior on violation, and \texttt{file:line} evidence.

\emph{Comparison agent.} $A_{\mathrm{cmp}}$ receives $\sigma(u)$ and both contracts and assigns each expected constraint $c \in K^{E}(u)$ one verdict $v(c)$. The verdict is \emph{satisfied} if the implementation enforces $c$ in the required phase, at the required assurance, on all paths. It is \emph{partial} if the implementation covers only some values, conditions, or paths; \emph{misplaced} if it covers $c$ in a later phase or at a weaker assurance level; \emph{missing} if nothing in $K^{I}(u)$ corresponds to $c$; and \emph{conflicting} if the implementation permits the behavior that $c$ forbids. Every other verdict is a \emph{gap}, with a counterexample and source evidence. Over a set of expected constraints $\mathcal{K}$, the gap rate is
\begin{equation}
  g(\mathcal{K}) = \frac{\left|\{\, c \in \mathcal{K} : v(c) \neq \textit{satisfied} \,\}\right|}{|\mathcal{K}|}.
  \label{eq:gaprate}
\end{equation}
Per call site, we also measure how much of the expected contract is satisfied:
\begin{equation}
  \rho(s) = \frac{\left|\{\, c \in K^{E}(s) : v(c) = \textit{satisfied} \,\}\right|}{|K^{E}(s)|},
  \label{eq:rho}
\end{equation}
where $K^{E}(s)$ is the union of the expected contracts of the units at $s$. Implemented constraints without an expected counterpart count only toward RQ2. Stage~II yields 579 responsibility units at the 496 call sites, with 5,017 implemented and 2,699 expected constraints.

\vspace{-4pt}
\subsection{Stage III: Exploitability and Fix Validation}
\label{sec:design-stage3}

Stage~III tests whether a contract gap matters at runtime and whether enforcing the missing constraint removes the risk. A gap matters only if a reply that violates the expected contract reaches a consumer that acts on it. Such a reply comes from an attacker who controls content that the model reads, such as a web page, a file, or a tool result~\cite{greshake2023indirect,liu2024formalizing}, or from a model that errs on its own. The attacker cannot change the code or the configuration of the system. The missing check is the same for both sources, so we study both and record each result's source. The authors perform all six steps manually, and Step~5 tests a real model against an oracle fixed in advance.

\input{tables/screening}

\emph{Step~1: Feasibility screening.} A gap can matter only through the code that consumes the output it constrains. We therefore link each gap $g$ to the consumption relations $R(g)$ of its output from Stage~I and audit each consumer in the code. The screen has two layers (Table~\ref{tab:screening}). Seven exclusion rules remove what clearly poses no threat, each with cited source lines. A remaining relation must then pass four conditions that form a \emph{threat chain}, the path along which a bad value travels from the defect to an effect at the consumer:
\begin{equation}
  \mathit{exploitable}(g) \iff \exists\, r \in R(g) : \neg E(r) \wedge C_1(r) \wedge C_2(r) \wedge C_3(r) \wedge C_4(r),
  \label{eq:exploitable}
\end{equation}
where $E(r)$ holds if an exclusion rule applies to $r$. A gap is thus excluded only when all its relations are, and one complete chain suffices to keep it.

\emph{Step~2: Gap attribution.} All gaps on an output share its chain, even when a gap is unrelated to the harmful value. For each exploitable gap, we therefore assume that the code enforces its constraint and recheck the chain. The gap is \emph{direct} if the chain breaks, \emph{partial} if a narrower chain remains, and \emph{shared} otherwise. We group direct and partial gaps by consumer and root cause into \emph{mechanisms}, because one fix at a consumer closes all gaps of its mechanism.

\emph{Step~3: Guided PoC.} For each mechanism, we pass the bad value that the chain predicts to the real downstream code, without a model. We count a downstream defense only if it is enabled by default and also stops a variant of the payload; a mechanism whose payload it fully stops leaves the exploitable set, and one stopped in part stays and is marked as partially neutralized. PoCs run each system's own code in sandboxed workspaces. \emph{Step~4: Severity grading.} We grade each PoC with CVSS v3.1~\cite{first2019cvss} from the chain evidence and label the harm with the OWASP LLM Top 10~\cite{owasp2025llm}.

\emph{Step~5: Trigger testing.} We test whether a model produces the bad value for all 58 mechanisms whose value arrives through a model reply, since selecting by severity would bias the result. For each, we try up to ten framings of the payload, stop at the first that reaches the consumer, and rerun it five times. A positive control sends the same request through a trusted channel, and we record the cause of every miss. All model calls go to one endpoint that serves \texttt{deepseek-v4-flash}, with each system's own requests and sampling parameters.

\emph{Step~6: Contract-guided fix.} For each triggered mechanism, we write the smallest patch that enforces the missing constraint in its expected phase, rerun the PoC with the original payload and one variant, and run the existing unit tests of the changed files when they run offline. A second round tries to close any residual risk under the same contract, and we report the strongest fix. For three label-only fixes, we compare labeled and unlabeled prompts on a real model.

\vspace{-4pt}
\subsection{Reliability and Validation}
\label{sec:design-reliability}

\input{tables/protocol}

LLM agents can err~\cite{zheng2023judging,chiang2023can}, so human reviewers check the output of every automated step, in full or on a sample, before it is used. Table~\ref{tab:protocol} summarizes the protocol; its icons mark steps done by static analysis (\faCogs), by an LLM agent in an isolated context (\faRobot), or manually (\faUser). Three authors of this paper serve as reviewers. Reviewers R1 and R2 label independently without seeing each other's labels, and R3 adjudicates every disagreement using the reasons of both. The reviewers refined the guidelines on six pilot systems, froze them, and then reviewed all 40 systems from scratch.

For Stages~I and~II, the reviewers made 17,428 judgments. Both reviewers check every item of the smaller steps. For the two largest steps, consumer tracing and implemented-contract extraction, both check a random sample of up to 40 items per system. Reviewers of the specifications and of the expected contracts work code-blind, in separate review batches that give access only to the specification and the expected contract; an audit of every such batch found no access to the source. They judge whether each specification is implementation-neutral, which fails if it names code identifiers, APIs, or vendors, conditions a duty on a code branch, or states an existing defect as a duty (AC1 0.92), and whether each expected constraint follows from the specification alone (AC1 0.95). Reviewers correct wrong items and add missed ones, and the final dataset includes these corrections. R3 resolved 604 disagreements in total. Coverage counts items before corrections, and comparison also covers implemented constraints without an expected counterpart.

Most automated results were judged correct, so the labels are highly skewed. Under such skew, Cohen's $\kappa$~\cite{cohen1960kappa} can be low even when agreement is near perfect~\cite{feinstein1990kappa}. We therefore report Gwet's AC1~\cite{gwet2008ac1}, which corrects for chance agreement using the category prevalence $\pi_k$ over $q$ categories,
\begin{equation}
  \mathrm{AC1} = \frac{p_a - p_e}{1 - p_e}, \qquad p_e = \frac{1}{q-1} \sum_{k=1}^{q} \pi_k (1 - \pi_k),
  \label{eq:ac1}
\end{equation}
where $p_a$ is the observed agreement, and we give $\kappa$ in parentheses. AC1 ranges from 0.91 to 0.99 across all steps, and it is 0.95 for the expected contracts and 0.96 for the verdicts.

\vspace{-4pt}
\section{Results}
\label{sec:results}

\subsection{RQ1: Which Logic Do Agentic Systems Delegate to LLM Calls, and Where Do the Outputs Flow?}
\label{sec:rq1}

\begin{figure*}[t]
  \centering
  \includegraphics[width=\textwidth]{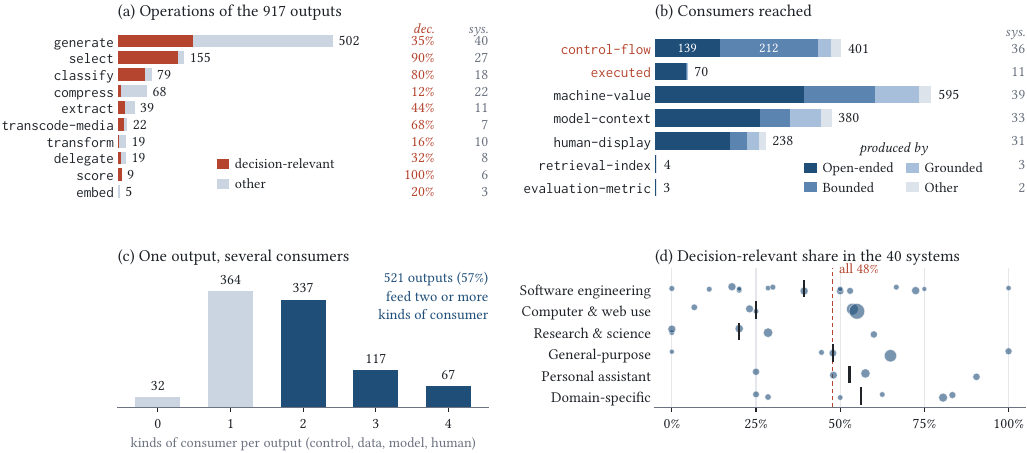}
  \caption{Operations and consumers of the 917 typed outputs. (a)~Outputs per operation. (b)~Outputs per consumer type. (c)~Consumer kinds per output. (d)~Decision-relevant share per system.}
  \label{fig:rq1-overview}
\end{figure*}

We answer RQ1 with the Stage~I profile of 917 typed outputs at 496 call sites and their 1,222 consumption relations (Figure~\ref{fig:rq1-overview}); an output counts once per consumer type.

\emph{Half of all outputs reach decisions, and much of this is free text.} Most call sites ask the model for open-ended text: \texttt{generate} accounts for 502 outputs (55\%) and occurs in all 40 systems. Yet 437 outputs (48\%) are decision-relevant. Bounded operations are decision-relevant almost by design (90\% of \texttt{select} and 80\% of \texttt{classify} outputs). The surprise is \texttt{generate}: 35\% of its outputs reach control flow or execution, which makes it the source of 175 of the 437 decision-relevant outputs (40\%). Of the 401 outputs that steer control flow, 139 are free text from \texttt{generate} (Figure~\ref{fig:rq1-overview}b), and 66 generated outputs are executed as code, 30 of which also steer control flow. Such outputs need a post-condition that decides which replies are accepted, and RQ3 shows that it is often weak.

\emph{Decision relevance varies by system more than by domain.} 36 of the 40 systems, from all six domains, have decision-relevant outputs, and the share varies more within a domain than across domains (Figure~\ref{fig:rq1-overview}d); software engineering agents range from 0\% to 100\%. How much a system lets its model decide depends on its own architecture.

\finding{1}{48\% of LLM outputs reach control flow or execution, and 40\% of these are free text.}

\input{listings/storm-check}

\emph{Most outputs never reach a person.} 71\% of the outputs, and 73\% of the decision-relevant ones, are used only by the program or by other models, so no user sees them before they take effect. 57\% feed two or more kinds of consumer (Figure~\ref{fig:rq1-overview}c), such as a tool call that is dispatched and also stored in the conversation that the next call reads. 227 generated outputs feed another model (Table~\ref{tab:roles}), and in six systems these hand-offs form chains of up to 12 steps (STORM). Counting outputs that reach a decision through such a chain raises the decision-relevant share from 48\% to 54\%, so an injected instruction early in a chain can shape decisions several calls later without any person in the loop.

\input{tables/roles}

\emph{Call sites play a small set of recurring roles.} Pairing an operation with a consumer type gives ten roles that cover 783 of the 917 outputs (85\%, Table~\ref{tab:roles}). Four of them steer the program, and two of these let unbounded text choose or become behavior: free-text steering (139 outputs, 20 systems), as when Aider adds every file named in a reply to the chat, and code execution (66 outputs, 9 systems).

\emph{One call site often plays several roles.} 29\% of the call sites produce more than one output, and 22\% produce both decision-relevant and other outputs, as in a TradingAgents call that selects tools to dispatch and writes a report for later agents. Such a call site needs a contract for each role. Figure~\ref{fig:storm-check} shows how thin the boundary between roles can be. STORM asks a model whether a user's topic is appropriate~\callout{1} and requests exactly ``Yes.'' or ``No.'', but tests the reply~\callout{2} only for a leading ``No''~\callout{3}. Any other reply, including a lowercase ``no'' or a refusal, approves the topic~\callout{4}: a free-text reply acts as a safety gate whose contract is a string prefix.

\finding{2}{71\% of LLM outputs are used only by the program or by other models, and 22\% of call sites serve decision-relevant and other roles at once.}

\vspace{-4pt}
\subsection{RQ2: How Do Agentic Systems Specify and Enforce the Contracts of Their LLM Calls?}
\label{sec:rq2}

\begin{figure*}[t]
  \centering
  \includegraphics[width=\textwidth]{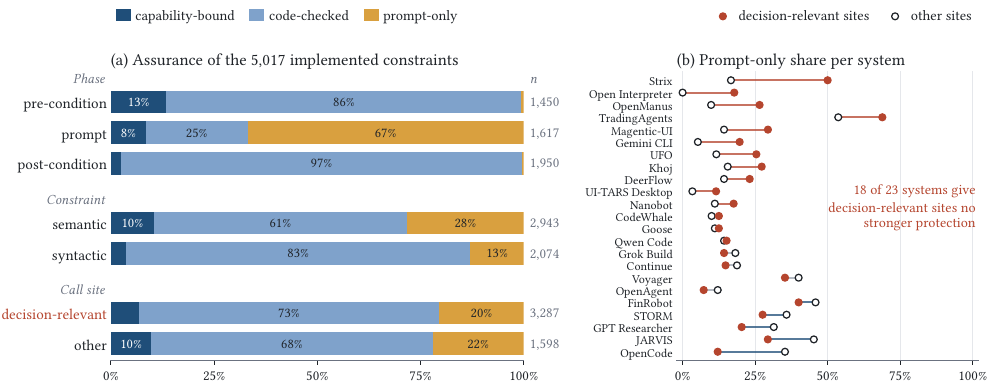}
  \caption{Assurance of implemented constraints. (a)~By phase, kind, call site. (b)~Prompt-only share per system.}
  \label{fig:rq2-enforcement}
\end{figure*}

We answer RQ2 with the 5,017 implemented constraints alone, 8.7 per unit on average. Figure~\ref{fig:rq2-enforcement}a breaks down their assurance levels (Eq.~\ref{eq:assurance}).

\emph{Programs constrain their calls, mostly in code.} 70.5\% of the implemented constraints are code-checked and 7.6\% are capability-bound, so 78\% are enforced by the program. Before the call, 86\% of constraints are code-checked and 13\% capability-bound, and after the call, 97\% are code-checked. Developers know that replies need checking; RQ3 asks whether their code checks the right things.

\emph{In the prompt, enforcement becomes a request.} In the prompt phase, 67\% of the constraints are prompt-only: the program asks the model to comply and never verifies that it did. Some cannot be checked by code, such as a request to reason carefully, but many can. Semantic constraints are prompt-only in 28\% of cases, but syntactic ones, which code can always decide, still are in 13\%. In total, 268 syntactic constraints in 27 systems are stated only in the prompt, for example limits on list length or the exact set of allowed names. The market analyst of TradingAgents asks for at most eight indicators with exact names, yet its code only checks that the reply contains a tool call. Each such constraint is decidable, cheap to check, and currently left to the model.

\finding{3}{Programs enforce 78\% of their constraints in code, but 67\% of prompt-phase constraints are only requested from the model, including 268 that code could decide.}

\input{tables/mechanisms}

\emph{Code checks come from a small set of mechanisms.} Table~\ref{tab:mechanisms} groups 3,748 of the 3,917 code-level constraints into 14 mechanisms; the other 169 fit none of them. After the call, programs mostly clean up and recover: extraction and normalization (557) and error handling and retry (479) are the most common mechanisms, while allow-list and range checks appear only 114 times. Before the call, programs check configuration and state far more often than they restrict the model's capabilities (207), and structured output, which makes a reply checkable~\cite{willard2023outlines}, is requested only 59 times. Current code keeps the pipeline running more than it keeps the model's authority small.

\emph{Decision-relevant call sites are not better protected.} If protection matched the stakes, decision-relevant call sites would rely less on the prompt and more on capabilities. They do not: at decision-relevant call sites, 20.5\% of constraints are prompt-only and 6.7\% capability-bound, against 22.1\% and 9.8\% at other call sites (Figure~\ref{fig:rq2-enforcement}a). Within systems, of the 23 systems with at least five constraints at both kinds of call site, 18 give their decision-relevant call sites no stronger protection (Figure~\ref{fig:rq2-enforcement}b), and in Strix and TradingAgents these call sites even rely on the prompt more than the others.

\finding{4}{Code checks favor recovery over validation, and decision-relevant call sites are not better protected than others in 18 of 23 systems.}

\vspace{-4pt}
\subsection{RQ3: How Adequate Are the Implemented Contracts, and Where Do They Fall Short?}
\label{sec:rq3}

\begin{figure*}[t]
  \centering
  \includegraphics[width=\textwidth]{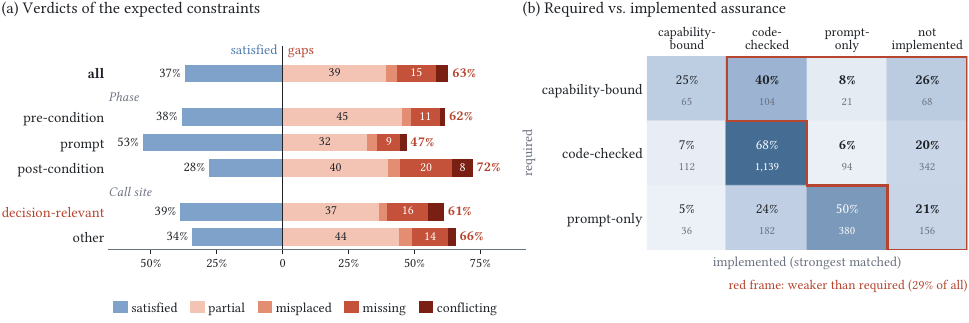}
  \caption{Verdicts of expected constraints. (a)~By phase and call site. (b)~Required vs.\ implemented assurance.}
  \label{fig:rq3-verdicts}
\end{figure*}

RQ3 uses the verdicts of 2,699 expected constraints, a median of four per unit (Figure~\ref{fig:rq3-verdicts}a).

\emph{Most expected constraints are unmet.} 63\% of the expected constraints have a gap (Eq.~\ref{eq:gaprate}). Partial coverage is the most common gap (39\%), followed by missing (15\%), conflicting (5\%), and misplaced constraints (4\%). In most gaps, the program does check the property, but only for some values, conditions, or paths. No system is free of gaps: the gap rate ranges from 33\% to 85\% per system (median 64\%), and 33 of the 40 systems miss more than half of their expected constraints.

\emph{Contracts break after the call, regardless of the stakes.} The gap rate is 72\% for post-conditions, 62\% for pre-conditions, and 47\% for prompt constraints. Post-conditions also hold most severe verdicts: 20\% are missing and 8\% conflicting. This inverts RQ2, where post-conditions looked best protected with 97\% code-checked: programs check their replies often, but cover less than the replies need. Decision-relevant call sites have a gap rate of 61\%, close to the 66\% of other call sites.

\emph{Implementations are weaker than required.} Figure~\ref{fig:rq3-verdicts}b compares the required assurance $\alpha^{E}(c)$ of each expected constraint with the strongest assurance $\alpha^{I}(c)$ among the implemented constraints matched to it, with $\alpha^{I}(c) = \bot$ if there are none. A constraint is enforced too weakly if $\alpha^{I}(c) \prec \alpha^{E}(c)$, which holds for 29\% of all expected constraints. Capability-bound constraints fare worst: only 25\% are enforced at that level, 40\% are downgraded to a code check, and 8\% to a prompt instruction. Developers thus strengthen the constraints that are easy to check and rarely build the capability boundaries that would make a violation impossible.

\finding{5}{63\% of the expected constraints are unmet, rising to 72\% for post-conditions, and 29\% are enforced more weakly than required.}

\input{tables/gap-patterns}

\begin{figure*}[t]
  \begin{minipage}[b]{0.41\textwidth}
    \centering
    \includegraphics[width=\linewidth]{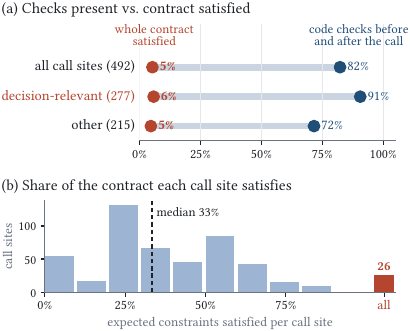}
    \caption{Checks present vs.\ contracts satisfied per call site.}
    \label{fig:rq3-adequacy}
  \end{minipage}\hfill
  \begin{minipage}[b]{0.57\textwidth}
    \input{listings/conseca-failopen}
  \end{minipage}
\end{figure*}

\emph{Having checks is not satisfying the contract.} A common audit asks whether each call site is guarded. By that measure, the systems look well protected: 82\% of call sites have code checks both before and after the call, and 91\% of the decision-relevant ones do (Figure~\ref{fig:rq3-adequacy}a). Measured against the expected contract (Eq.~\ref{eq:rho}), the picture changes. Only 26 of the 492 call sites with judged expected constraints (5.3\%) reach $\rho(s) = 1$, and the median call site satisfies one third of its contract (Figure~\ref{fig:rq3-adequacy}b). Decision-relevant call sites do no better (5.8\%). The distance between 82\% and 5.3\% is the central result of this study. Programs check the form of a reply, such as whether it parses, while the contract also requires its meaning, such as whether a proposed action is among those offered. A metric that counts guards would call these systems safe; a contract-level analysis shows that almost none of their call sites are.

\finding{6}{82\% of call sites check in code before and after the call, yet only 5.3\% satisfy their whole expected contract, and the median call site satisfies one third.}

\emph{Ten patterns explain almost all gaps.} Coding each gap by the reason it exists yields ten patterns that cover 1,685 of the 1,697 gaps (99\%, Table~\ref{tab:gap-patterns}). Each recurs in 12 to 34 systems, so the gaps are systematic, and half of them (49\%) arise after the call. The most consequential ones concentrate where the stakes are. 73\% of the unvalidated actions (138) occur at decision-relevant call sites, and fail-open handling (146) accounts for 57 of the 130 conflicting verdicts (44\%) by substituting a default or passing a failure on as a normal result. Path-dependent checks (101) hold on the main path only, as in SWE-agent, which enforces one action per reply only for function calls.

\emph{Fail-open handling defeats layered defenses.} Figure~\ref{fig:conseca} shows fail-open handling at a safety gate. The opt-in policy enforcer of Gemini CLI asks a model whether a proposed tool call complies with a security policy, and the model must answer allow, deny, or ask the user~\callout{1}. The code judges the call against an empty policy when the tool has none~\callout{2}, and returns \texttt{ALLOW} when the reply is empty~\callout{3}, does not parse~\callout{4}, or the request fails~\callout{5}. Five of the six return paths allow the call, and only a reply that parses can deny it. Downstream, two guards fail closed when a checker throws, but the enforcer converts every failure into an ordinary \texttt{ALLOW} first, so neither ever fires: the model is asked the right question, and the program does not enforce the answer.

\finding{7}{Ten recurring patterns explain 99\% of the gaps, and fail-open handling alone accounts for 44\% of the conflicting verdicts.}

\vspace{-4pt}
\subsection{RQ4: What Are the Runtime Consequences of Contract Gaps, and Does Enforcing the Missing Constraint Prevent Them?}
\label{sec:rq4}

\begin{table*}[t]
  \caption{Gaps and mechanisms through the six steps of Stage~III.}
  \label{tab:rq4-funnel}
  \centering
  \scriptsize
  \setlength{\tabcolsep}{2.5pt}
  \renewcommand{\arraystretch}{1.0}
  \begin{tabular*}{\textwidth}{@{\extracolsep{\fill}}l r rrrrrrrrr rrrr rrrr@{}}
    \toprule
     & & \multicolumn{9}{c}{\emph{Step 1: Feasibility screening}} & \multicolumn{4}{c}{\emph{Step 2: Gap attribution}} & \multicolumn{4}{c}{\emph{Steps 3 to 6}} \\
    \cmidrule(lr){3-11}\cmidrule(lr){12-15}\cmidrule(l){16-19}
     & RQ3 & E1 & E2 & E3 & E4 & E5 & E6 & E7 & No chain & Expl. & Direct & Partial & Shared & On chain & PoC & Trig. & Patched & Blocked \\
    \midrule
    Gaps & 1,697 & 716 & 30 & 165 & 138 & 84 & 132 & 137 & 5 & 290 & 48 & 66 & 176 & 114 & \multicolumn{1}{c}{--} & \multicolumn{1}{c}{--} & 68 & 32 \\
    Mech. & \multicolumn{1}{c}{--} & \multicolumn{1}{c}{--} & \multicolumn{1}{c}{--} & \multicolumn{1}{c}{--} & \multicolumn{1}{c}{--} & \multicolumn{1}{c}{--} & \multicolumn{1}{c}{--} & \multicolumn{1}{c}{--} & \multicolumn{1}{c}{--} & \multicolumn{1}{c}{--} & \multicolumn{1}{c}{--} & \multicolumn{1}{c}{--} & \multicolumn{1}{c}{--} & 60 & 60 & 38 & 38 & 23 \\
    \bottomrule
  \end{tabular*}
\end{table*}

\emph{Most gaps never reach a consumer.} Table~\ref{tab:rq4-funnel} follows the 1,697 gaps of RQ3 through the six steps of Stage~III. Screening excludes 1,407 of the 1,697 gaps (83\%). The largest group, 716 gaps, arises before the reply exists and changes nothing that a consumer does. Independent checks and working validators stop 303 more, 84 need a non-default configuration or sit in dead code, and 137 concern values that never reach security-relevant logic.

\emph{Few gaps lie on the threat chain.} 290 gaps (17\%) are technically exploitable, and for 114 of them (39\%) the missing constraint lies on the threat chain: enforcing it breaks the chain for 48 gaps and narrows it for 66. The other 176 miss a property, such as a non-empty reply, that plays no role in the attack. Unvalidated actions lie on the chain most often (23 of 26, Table~\ref{tab:rq4-patterns}), and path-dependent checks are mostly partial (23 of 30), because the attack takes a path that the check does not cover.

\begin{wraptable}{r}{0.5\textwidth}
  \vspace{-\intextsep}
  \caption{Exploitable gaps by pattern.}
  \label{tab:rq4-patterns}
  \centering
  \scriptsize
  \setlength{\tabcolsep}{3pt}
  \renewcommand{\arraystretch}{1.0}
  \begin{tabularx}{\linewidth}{@{}X r r r r@{}}
    \toprule
    Pattern & Gaps & Exploitable & Direct & Partial \\
    \midrule
    Unchecked call inputs & 330 & 7 & 3 & 0 \\
    Over-broad authority & 90 & 15 & 7 & 3 \\
    Underspecified prompt & 242 & 9 & 0 & 8 \\
    Faulty request assembly & 93 & 12 & 1 & 11 \\
    Partial format check & 249 & 72 & 6 & 5 \\
    Unchecked content & 253 & 81 & 11 & 9 \\
    Fail-open handling & 146 & 23 & 0 & 1 \\
    Unvalidated action & 138 & 26 & 19 & 4 \\
    Prompt-only check & 43 & 14 & 1 & 1 \\
    Path-dependent check & 101 & 30 & 0 & 23 \\
    Other & 12 & 1 & 0 & 1 \\
    \midrule
    Total & 1,697 & 290 & 48 & 66 \\
    \bottomrule
  \end{tabularx}
  \vspace{-\intextsep}
\end{wraptable}

\emph{The exploitable mechanisms are severe.} The 60 mechanisms behind the 114 gaps score 9 critical, 21 high, 26 medium, and 4 low on CVSS, and the critical ones sit at control-flow, machine-value, and execution consumers, so severity follows the consumer that RQ1 identifies. By OWASP category~\cite{owasp2025llm}, 17 are excessive agency and 15 improper output handling. 38 need an attacker, 5 only a model error, and 17 either.

\emph{Most exploitable mechanisms trigger in practice.} 38 of the 60 mechanisms trigger: 36 against a real model, and two deterministically, because their bad value needs no model call. Table~\ref{tab:rq4-triggers} shows seven of the 58 model tests across the range of outcomes. Some need no attacker: JARVIS chose a tool outside its candidate pool in all five runs, because the model invented identifiers~\cite{ji2023hallucination} that exist in the real tool catalog. In skyvern, the policy engine denies the click in every run, but the dispatch code never reads the decision.

\begin{table}[t]
  \caption{Seven trigger tests against a real model, grouped by outcome.}
  \label{tab:rq4-triggers}
  \centering
  \scriptsize
  \setlength{\tabcolsep}{3pt}
  \renewcommand{\arraystretch}{1.0}
  \begin{tabularx}{\linewidth}{@{}l >{\raggedright\arraybackslash}X l r >{\raggedright\arraybackslash}p{0.33\linewidth}@{}}
    \toprule
    System & Mechanism & Source & Hits & Causes of misses \\
    \midrule
    \multicolumn{5}{@{}l}{\emph{Triggered in every run}} \\
    \cmidrule{1-5}
    AgentLaboratory & bare regex passes model text to \texttt{exec()} & attacker & 5/5 & \\
    JARVIS & picks a tool from an empty candidate pool & model error & 5/5 & \\
    gpt-researcher & chat renders report text as HTML & attacker & 5/5 & \\
    \midrule
    \multicolumn{5}{@{}l}{\emph{Triggered in some runs}} \\
    \cmidrule{1-5}
    skyvern & browser action policy only observes & attacker & 4/5 & cross-domain link flagged (1) \\
    Codewhale & \texttt{repl} fence found by substring (Figure~\ref{fig:motivating}) & attacker, model & 1/10 & parser unmet (6), refused (2), off-task (1) \\
    \midrule
    \multicolumn{5}{@{}l}{\emph{Never triggered}} \\
    \cmidrule{1-5}
    OpenManus & \texttt{python\_execute} runs model-written code & attacker & 0/10 & injection recognized (10) \\
    khoj & tool dispatch skips the authorization check & attacker & 0/10 & refused (4), ignored (4), wrong format (2) \\
    \bottomrule
  \end{tabularx}
\end{table}

\emph{Misses rarely come from a working check.} A low trigger rate can hide luck. Of the 22 mechanisms that never triggered, the model or a serving-layer safety mechanism declined the value for 15, the model made no mistake for three non-adversarial ones, and four could not be tested to completion.

\finding{8}{17\% of the gaps are technically exploitable, the missing constraint lies on the threat chain for 7\% (114 gaps), and 38 of the 60 mechanisms behind them trigger.}

\emph{A small code fix stops every exploit, and 23 of 38 completely.} We patched all 38 triggered mechanisms, which span 19 systems and 68 gaps, with patches of 1 to 366 lines. On the patched copies, the original payload fails for all 38 and the variant for 35. 23 mechanisms (32 gaps) are blocked completely, with no variant or sibling path left. The other 15 stay partial because they need semantic judgment that a pattern cannot give, as in the goal loop of deer-flow, rely on an in-process sandbox without a hard boundary, as in Voyager, or need infrastructure that a patch cannot add; for three, the residual case lies outside the constraint or is a risk the system accepts on purpose. Existing unit tests run offline for 19 patched mechanisms and all pass. Fixes that only label untrusted content are inconsistent: of three mechanisms, the label removed the effect for one, left another unchanged, and had nothing to remove for the third, whose baseline was already zero.

\finding{9}{A minimal code fix that enforces the expected constraint stops the original exploit for all 38 mechanisms and blocks 23 completely, while fixes that only label untrusted content work for some mechanisms and not others.}

\emph{Case study: Codewhale.} For the mechanism of Figure~\ref{fig:motivating}, Stage~II marks both expected constraints as partial, and six of nine misses lacked the closing fence that the parser needs. A real model never ran the attacker's code, but because the detector matches the marker anywhere, one refusal that mentioned the marker made the loop run two fragments of its explanation as Python: the gap has two triggers, an attacker and a model that mentions the fence.

\begin{figure*}[t]
  \begin{minipage}[t]{0.49\textwidth}
    \centering
    \scalebox{1.02}{\input{figures/case-gptresearcher}}\\[2pt]
    {\small (a) gpt-researcher}
  \end{minipage}\hfill
  \begin{minipage}[t]{0.49\textwidth}
    \centering
    \scalebox{1.02}{\input{figures/case-khoj}}\\[2pt]
    {\small (b) khoj}
  \end{minipage}
  \caption{Two mechanisms of Table~\ref{tab:rq4-triggers}. (a)~A render path reaches the browser with no sanitizer. (b)~An unscoped opt-in reaches an internal address, with no injection needed.}
  \label{fig:rq4-cases}
\end{figure*}
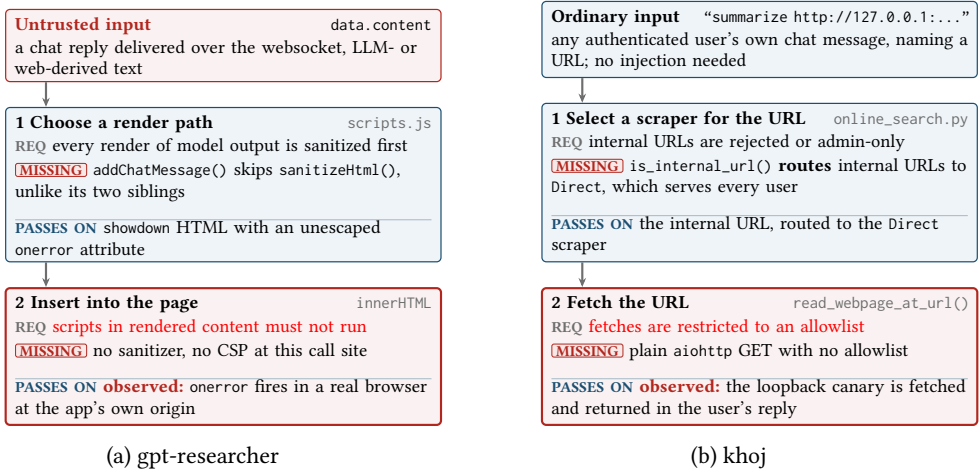

\emph{Case study: gpt-researcher.} In Figure~\ref{fig:rq4-cases}a, the chat panel renders replies into \texttt{innerHTML} without the sanitizer that an earlier cross-site scripting fix added to two sibling renderers. A blunt request to echo a scripted payload mostly fails, but the same payload in one cell of an ordinary markdown table runs in a real browser in all five runs.

\emph{Case study: khoj.} In Figure~\ref{fig:rq4-cases}b, the \texttt{Direct} scraper fetches internal and loopback URLs with no allowlist. Its commit warns that it must stay off outside single-user deployments, yet once an admin enables it, every user's chat uses it, and an unprivileged account fetched a loopback page in ordinary chat. No injection is needed: the gap lies in the scope of an intentional feature.

\emph{Disclosure.} We have reported all 38 triggered mechanisms to the maintainers through private security channels, together with our patches; the disclosure status of each system is on our website.
\vspace{-4pt}
\section{Discussion and Implications}
\label{sec:discussion}

When logic moves into an LLM call, its specification moves into the program around the call site. Programs re-establish part of it in code but leave much to the prompt (Finding~3), check the form of a reply more than its meaning (Findings~5 and~6), break their checks on failure paths~\cite{yuan2014simple} (Finding~7), and do not protect decisions more than other outputs (Finding~4); on a threat chain, such a gap becomes a runtime failure (Findings~8 and~9).

\textbf{Implications for agent developers.} (1)~\emph{Restate the contract of delegated logic, per role.} Our admission criteria (Section~\ref{sec:design-stage2}) give a checklist, and since 22\% of call sites serve several roles (Finding~2), each role needs its own contract. (2)~\emph{Prioritize by consumer.} Only 7\% of the gaps lie on a threat chain (Finding~8), and all nine critical mechanisms sit at control-flow, machine-value, or execution consumers, so the consumer type (Table~\ref{tab:taxonomy}) is a cheap triage rule. (3)~\emph{Check in code what code can decide.} Programs state 268 syntactic constraints only in the prompt (Finding~3). Code fixes stopped the original exploit for all 38 mechanisms, whereas labeling untrusted content in the prompt removed the effect for only one of the two mechanisms that showed one (Finding~9). (4)~\emph{Validate the meaning of a reply.} Post-conditions are 97\% code-checked yet have the highest gap rate (72\%, Finding~5): programs check whether a reply parses more than whether it is admissible. Unvalidated actions lie on a threat chain most often (23 of 26 exploitable gaps) and need no attacker, as JARVIS shows. (5)~\emph{Fail closed on every path and honor every verdict.} Gemini CLI turns every failure of its policy check into an approval (Figure~\ref{fig:conseca}), skyvern computes a denial that its dispatch code never reads, and an earlier fix in gpt-researcher sanitized two renderers and missed the third (Figure~\ref{fig:rq4-cases}a). Enforcing the contract closes such sibling paths together~\cite{saltzer1975protection}. (6)~\emph{Do not count on the model as the defense.} For 15 of the 22 mechanisms that never triggered, the model or the serving layer declined the value. As the model is part of the implementation, replacing it changes behavior without changing code; constraints enforced in code hold across models, and refusals may not.

\textbf{Implications for framework maintainers.} Safe defaults at the boundaries that frameworks provide help every agent built on them: tool dispatch should reject names and arguments outside the declared tools, parsers should raise on a failure and never fall back to free text, and guards should return a typed denial. Only 25\% of the constraints that require a capability boundary have one (Section~\ref{sec:rq3}); scoping tools to each call would make least privilege~\cite{saltzer1975protection} the default.

\textbf{Implications for researchers and tool builders.} (1)~\emph{Measure adequacy, not presence.} Guard counts are a poor proxy for protection: 82\% of call sites are checked before and after the call, yet 5.3\% satisfy their whole contract (Finding~6). (2)~\emph{Use contracts as test oracles.} Testing cannot establish what an LLM call guarantees, partly because it lacks an oracle~\cite{barr2015oracle}; a blindly derived expected contract supplies one for tests and runtime monitors~\cite{leucker2009brief}. (3)~\emph{Build consumer-aware detectors.} Analyzers that treat model replies as untrusted input~\cite{arzt2014flowdroid} could flag signatures such as a handler that returns approval or a verdict tested by a string prefix (Figure~\ref{fig:storm-check}), filtered by our screening rules (Table~\ref{tab:screening}). (4)~\emph{Evaluate per call site and per model.} Trigger outcomes depend on the model, so exploitability should be reported per model.

\vspace{-4pt}
\section{Threats to Validity}
\label{sec:threats}

\emph{Construct validity.} The expected contract is a normative reconstruction, so other analysts might admit slightly different constraints. The admission criteria, the code-blind review of every expected constraint (AC1 0.95), and the runtime tests of RQ4 keep it close to what a responsibility needs. A further concern is that $\sigma$ is written with the code in view and could carry implementation choices into $K^{E}$; besides the neutrality filter and the code-blind review, the data argue against such leakage. Code-blind reviewers flagged 48 specifications as not neutral; excluding their units changes the gap rate by less than one percentage point, and these units have no fewer gaps than the others, so leakage, where it occurred, did not make the expected contract track the code. Moreover, 15\% of the expected constraints have no counterpart in the code at all, which a contract copied from the implementation could not produce. Stage~I categories involve judgment, constrained by boundary rules and review (AC1 $\geq$ 0.91 for all Stage~I steps).

\emph{Internal validity.} LLM agents can err, so we isolate their contexts, require source evidence for every claim, and review every step of Stages~I and~II. Static identification can miss call sites reached through dynamic dispatch or reflection; reviewers examined all 1,226 candidates and added the missed ones. In Stage~III, a wrong manual exclusion can hide an exploit, so every exclusion cites source lines, and excluding outputs without consumers makes 290 a lower bound.

\emph{External validity.} Our 40 open-source systems span six domains and four languages and are analyzed at one commit each; closed-source agents and agent frameworks may differ. Every system has gaps, and the gap rate does not depend on system size (Pearson's $r = -0.07$). Trigger tests use one substitute model and at most ten framings, so they show that a mechanism can trigger, not how often it does under each system's default model.

\vspace{-4pt}
\section{Related Work}
\label{sec:related}

\textbf{Empirical studies of agents and LLM-based software.} Software engineering research has studied machine learning components inside programs, from technical debt~\cite{sculley2015hidden} and engineering practice~\cite{amershi2019se4ml} to the bugs of deep learning systems~\cite{islam2019dlbugs,humbatova2020taxonomy}. For LLMs, studies survey their use in software engineering~\cite{hou2024llm4se}, classify LLM-integrated applications~\cite{weber2024taxonomy}, and mine how developers write prompts~\cite{liang2025prompts,tafreshipour2025prompting}. Agents build on tool use and multi-agent collaboration~\cite{yao2023react,schick2023toolformer,wu2024autogen,hong2024metagpt}, are evaluated by task success~\cite{jimenez2024swebench,liu2024agentbench,yao2024taubench}, and are studied for their designs~\cite{xia2025agentless} and failures~\cite{cemri2025mast}. These works analyze the model, the prompt, or the whole agent; we analyze the call site and its contract.

\textbf{Security of LLM-based applications.} Prompt injection lets content that a model reads override its developer's instructions~\cite{perez2022ignore,greshake2023indirect,liu2023promptinjection,liu2024formalizing}, and jailbreaks bypass safety training~\cite{wei2023jailbroken}. Benchmarks measure how agents fall to injected content~\cite{zhan2024injecagent,debenedetti2024agentdojo,zhang2025asb} and to their own errors~\cite{ruan2024toolemu}; plugins widen the attack surface~\cite{iqbal2024plugins}, and generated code carries vulnerabilities~\cite{pearce2022asleep}. Defenses separate instructions from data~\cite{chen2025struq,hines2024spotlighting}, isolate components~\cite{wu2025isolategpt}, constrain data flow~\cite{debenedetti2025camel}, filter inputs and outputs~\cite{rebedea2023nemo,inan2023llamaguard}, or restrict generation~\cite{beurer2023lmql,khattab2024dspy,willard2023outlines}. These works ask whether a model or an agent resists attacks; we ask which checks the surrounding programs implement and which are missing.

\textbf{Contracts and specification mining.} Contracts go back to program verification~\cite{floyd1967assigning,hoare1969axiomatic} and design by contract~\cite{meyer1992applying}; JML and Spec\# make them checkable~\cite{leavens2006jml,barnett2005spec}, though most contracts in practice are simple~\cite{estler2014contracts}. Specification mining infers contracts from behavior~\cite{ammons2002mining,ernst2007daikon}, API-misuse detectors check usage constraints~\cite{amann2019api}, machine learning APIs have their own contracts~\cite{khairunnesa2023contracts}, and LLMs generate specifications and postconditions~\cite{ma2025specgen,endres2024nl2postcond}. We derive expected contracts without the code, which makes omissions measurable for a probabilistic callee that its input can steer.

\vspace{-4pt}
\section{Conclusion}
\label{sec:conclusion}

When code becomes an LLM call, its specification must be enforced around the call site. In 496 call sites of 40 agentic systems, 63\% of the expected constraints are unmet and only 5.3\% of call sites satisfy their whole contract, yet a small code fix that enforces the missing constraint stops all 38 triggered exploits. We hope that our contract view, gap patterns, and dataset help developers build call sites that check what they need.

\bibliographystyle{ACM-Reference-Format}
\bibliography{reference}

\end{document}

%% file: figures/motivating-example.tex
\begingroup
\definecolor{advred}{HTML}{B3261E}
\definecolor{gapmiss}{HTML}{B3261E}
\definecolor{gappart}{HTML}{A8620A}
\definecolor{sysblue}{HTML}{1B4F72}
\definecolor{sysbg}{HTML}{EFF4F9}
\definecolor{mutegray}{HTML}{6E6E6E}
\newcommand{\cd}[1]{{\ttfamily\fontsize{6.2}{7.2}\selectfont #1}}
\newcommand{\reqlbl}[1]{{\fontsize{5.6}{5.6}\selectfont\bfseries\color{mutegray}#1}}
\newcommand{\chip}[2]{%
  \tikz[baseline=(c.base)]{\node[inner xsep=1.6pt, inner ysep=0.5pt,
    rounded corners=0.8pt, fill=#1!10, draw=#1, line width=0.3pt, text=#1,
    font=\fontsize{5.6}{5.6}\selectfont\bfseries] (c) {#2};}}
\newcommand{\passes}[1]{%
  \\[0.7pt]{\color{sysblue!40}\rule{\linewidth}{0.25pt}}\\[-3.3pt]
  {\color{sysblue}\fontsize{5.6}{5.6}\selectfont\bfseries PASSES ON}~#1}
\begin{tikzpicture}[
  font=\fontsize{6.8}{8.0}\selectfont,
  stage/.style={draw=sysblue, fill=sysbg, rounded corners=1.6pt, line width=0.45pt,
                anchor=north, inner xsep=3.5pt, inner ysep=2.6pt,
                text width=64mm, align=left},
  src/.style={stage, draw=advred, fill=advred!6},
  hit/.style={stage, draw=advred, fill=advred!6, line width=0.8pt},
  flow/.style={-{Stealth[length=3.4pt,width=2.8pt]}, line width=0.8pt, draw=mutegray},
]
\def\bx{0mm}
\def\fx{-28mm}

\node[src] (cfg) at (\bx,0) {%
  \textcolor{advred}{\bfseries Untrusted input}\hfill\cd{dev-setup-notes.txt}\,\textperiodcentered\,a file in the workspace\\[0.6pt]
  a note whose text asks the assistant to answer in a \cd{repl} code fence \emph{(paraphrased)}%
};

\node[stage, below=3.2mm of cfg.south, anchor=north] (s1) {%
  \textbf{1\,\,Build the request}%
  \hfill{\color{mutegray}\cd{turn\_loop.rs}\,\textperiodcentered\,pre-call}\\[0.9pt]
  \reqlbl{REQ}~execution must be restricted to the tools advertised for this session\\[0.9pt]
  \chip{gappart}{PARTIAL}~the \cd{repl} fence is a second way to run code that is not an advertised tool
  \passes{the file text as a tool result $\Rightarrow$ a reply that contains a \cd{repl} fence}%
};

\node[stage, below=3.2mm of s1.south, anchor=north] (s2) {%
  \textbf{2\,\,Detect code in the reply}%
  \hfill{\color{mutegray}\cd{sandbox.rs:9--11}\,\textperiodcentered\,post-call}\\[0.9pt]
  \reqlbl{REQ}~only completed tool-call blocks are executed\\[0.9pt]
  \chip{gappart}{PARTIAL}~any reply that contains the marker counts as code: \cd{text.contains("\textasciigrave\textasciigrave\textasciigrave repl")}
  \passes{the text after the marker, up to the next closing fence}%
};

\node[hit, below=3.2mm of s2.south, anchor=north] (s3) {%
  \textbf{3\,\,Execute the extracted code}%
  \hfill{\color{mutegray}\cd{runtime.rs:1018}\,\textperiodcentered\,post-call}\\[0.9pt]
  \reqlbl{REQ}~code runs only as an approved tool call\\[0.9pt]
  \chip{gapmiss}{MISSING}~\cd{exec()} in a \cd{python3} subprocess: no approval, no sandbox (the docs say ``sandboxed'')
  \passes{\textcolor{advred}{\bfseries observed:} a fixed fence runs as real Python; a real model's prose about the marker raised a \cd{SyntaxError}}%
};

\foreach \a/\b in {cfg/s1, s1/s2, s2/s3}
  \draw[flow] (\fx,0 |- \a.south) -- (\fx,0 |- \b.north);

\end{tikzpicture}
\endgroup

%% file: tables/subjects.tex
\begin{table*}[t]
  \caption{Subject systems, grouped by task domain and sorted by GitHub stars.}
  \label{tab:subjects}
  \centering
  \scriptsize
  \setlength{\tabcolsep}{0.8pt}
  \renewcommand{\arraystretch}{1.0}
\begin{tabular*}{\textwidth}{@{}l@{\extracolsep{\fill}}l@{}}
\toprule
\begin{tabular*}{0.495\textwidth}[t]{@{}c@{\hspace{2pt}}l@{\extracolsep{\fill}}c r r c r r r@{}}
& \multicolumn{6}{c}{\textbf{Subject}} & \multicolumn{2}{c}{\textbf{Analyzed}} \\
\cmidrule(lr){2-7}\cmidrule(l){8-9}
& System & Lang & \staricon & Contr. & Since & KLOC & Cand. & Sites \\
\midrule
 & OpenCode & \langicon{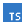} & 210k & 1,016 & 2025 & 450 & 16 & 9 \\
 & Codex & \langicon{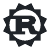} & 126k & 625 & 2025 & 894 & 13 & 2 \\
 & Gemini CLI & \langicon{typescript} & 107k & 698 & 2025 & 249 & 52 & 7 \\
 & OpenHands\,\cite{wang2025openhands} & \langicon{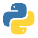} & 89k & 554 & 2024 & 95 & 29 & 6 \\
 & Cline & \langicon{typescript} & 69k & 328 & 2024 & 202 & 10 & 2 \\
 & Open Interpreter & \langicon{rust} & 68k & 616 & 2023 & 894 & 16 & 3 \\
 & Aider & \langicon{python} & 49k & 182 & 2023 & 18 & 7 & 3 \\
 & CodeWhale & \langicon{rust} & 41k & 229 & 2026 & 951 & 36 & 26 \\
 & Continue & \langicon{typescript} & 36k & 507 & 2023 & 201 & 32 & 9 \\
 & Qwen Code & \langicon{typescript} & 28k & 576 & 2025 & 1,532 & 50 & 13 \\
 & Grok Build & \langicon{rust} & 27k & 1 & 2026 & 1,240 & 24 & 2 \\
 & Prime Agent & \langicon{typescript} & 21k & 225 & 2026 & 174 & 27 & 10 \\
 & SWE-agent\,\cite{yang2024sweagent} & \langicon{python} & 20k & 101 & 2024 & 11 & 10 & 1 \\
 & Devika & \langicon{python} & 20k & 47 & 2024 & 3.6 & 21 & 7 \\
\multirow{-15}{*}{\grouplabel{Software\\Engineering}} & Plandex & \langicon{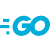} & 16k & 27 & 2023 & 60 & 11 & 1 \\
\cmidrule{1-9}
 & UI-TARS Desktop\,\cite{qin2025uitars} & \langicon{typescript} & 39k & 49 & 2025 & 123 & 31 & 26 \\
 & Skyvern & \langicon{python} & 23k & 95 & 2024 & 606 & 14 & 15 \\
 & Agent S\,\cite{agashe2025agent} & \langicon{python} & 12k & 27 & 2024 & 17 & 49 & 33 \\
 & Magentic-UI & \langicon{python} & 10k & 35 & 2025 & 37 & 4 & 4 \\
\multirow{-5}{*}{\grouplabel{Computer \&\\Web Use}} & UFO\,\cite{zhang2024ufo} & \langicon{python} & 9.8k & 30 & 2024 & 73 & 30 & 13 \\

\end{tabular*} &
\begin{tabular*}{0.499\textwidth}[t]{@{}c@{\hspace{2pt}}l@{\extracolsep{\fill}}c r r c r r r@{}}
& \multicolumn{6}{c}{\textbf{Subject}} & \multicolumn{2}{c}{\textbf{Analyzed}} \\
\cmidrule(lr){2-7}\cmidrule(l){8-9}
& System & Lang & \staricon & Contr. & Since & KLOC & Cand. & Sites \\
\midrule
 & DeerFlow & \langicon{python} & 83k & 467 & 2025 & 176 & 98 & 18 \\
 & STORM\,\cite{shao2024storm} & \langicon{python} & 31k & 26 & 2024 & 10 & 46 & 40 \\
 & GPT Researcher & \langicon{python} & 30k & 271 & 2023 & 30 & 26 & 6 \\
 & RD-Agent & \langicon{python} & 15k & 40 & 2024 & 49 & 90 & 8 \\
\multirow{-5}{*}{\grouplabel{Research \&\\Science}} & Agent Laboratory\,\cite{schmidgall2025agentlab} & \langicon{python} & 5.9k & 10 & 2025 & 3.6 & 43 & 27 \\
\cmidrule{1-9}
 & OpenManus & \langicon{python} & 58k & 62 & 2025 & 10 & 10 & 7 \\
 & Goose & \langicon{rust} & 55k & 642 & 2024 & 309 & 36 & 13 \\
 & AgenticSeek & \langicon{python} & 27k & 51 & 2025 & 7.4 & 28 & 18 \\
 & JARVIS\,\cite{shen2023hugginggpt} & \langicon{python} & 25k & 23 & 2023 & 5.9 & 60 & 57 \\
\multirow{-5}{*}{\grouplabel{General-\\Purpose}} & XAgent & \langicon{python} & 8.6k & 38 & 2023 & 17 & 12 & 7 \\
\cmidrule{1-9}
 & OpenClaw & \langicon{typescript} & 390k & 3,416 & 2025 & 4,186 & 71 & 13 \\
 & Nanobot & \langicon{python} & 49k & 457 & 2026 & 174 & 53 & 21 \\
 & Khoj & \langicon{python} & 37k & 76 & 2021 & 57 & 46 & 13 \\
\multirow{-4}{*}{\grouplabel{Personal\\Assistant}} & OpenAgent & \langicon{go} & 5.7k & 65 & 2020 & 110 & 50 & 18 \\
\cmidrule{1-9}
 & TradingAgents\,\cite{xiao2024tradingagents} & \langicon{python} & 108k & 21 & 2024 & 8.5 & 22 & 13 \\
 & Strix & \langicon{python} & 64k & 71 & 2025 & 51 & 9 & 6 \\
 & PentAGI & \langicon{go} & 25k & 17 & 2025 & 250 & 13 & 2 \\
 & Data Formulator\,\cite{wang2024dataformulator} & \langicon{python} & 17k & 37 & 2024 & 88 & 16 & 2 \\
 & FinRobot\,\cite{yang2024finrobot} & \langicon{python} & 8.1k & 7 & 2024 & 19 & 8 & 8 \\
\multirow{-6}{*}{\grouplabel{Domain-\\Specific}} & Voyager\,\cite{wang2024voyager} & \langicon{python} & 7.2k & 13 & 2023 & 8.5 & 7 & 7 \\

\end{tabular*} \\
\bottomrule
\end{tabular*}
\end{table*}

%% file: tables/taxonomy.tex
\begin{table*}[t]
  \caption{Classification scheme for the outputs of LLM calls.}
  \label{tab:taxonomy}
  \centering
  \scriptsize
  \setlength{\tabcolsep}{2pt}
  \renewcommand{\arraystretch}{1.0}
  \begin{tabular*}{\textwidth}{@{\extracolsep{\fill}}l l l l l@{}}
    \toprule
    \multicolumn{5}{@{}l}{\textbf{Operation}: what the model is asked to do} \\
    \midrule
    \emph{Group} & \emph{Category} & \emph{Definition} & \emph{Boundary rule} & \emph{Example} \\
    \cmidrule{1-5}
    Open-ended & \texttt{generate} & New free-form content & not already in the input (vs.\ extract) & write a report or code \\
    \cmidrule{1-5}
    Grounded & \texttt{compress} & Lossy condensation of the input & drops information (vs.\ transform) & summarize a conversation \\
             & \texttt{extract} & Structured values from the input & copied, not created (vs.\ generate) & read fields from a page \\
             & \texttt{transform} & Meaning-preserving rewrite & keeps meaning (vs.\ compress) & reformulate a query \\
    \cmidrule{1-5}
    Bounded & \texttt{select} & Next action, tool, or resource & picks what to do next (vs.\ classify) & pick a tool and arguments \\
             & \texttt{classify} & Label from a bounded set & discrete values (vs.\ score) & judge if a step succeeded \\
             & \texttt{score} & Ordinal or continuous value & ordered values (vs.\ classify) & rate a candidate answer \\
    \cmidrule{1-5}
    Orchestration & \texttt{delegate} & Whole task to an agent loop & callee is a multi-turn agent & run a sub-agent \\
    \cmidrule{1-5}
    Non-text & \texttt{transcode-media} & Non-text media & output is not text & synthesize speech \\
             & \texttt{embed} & Numeric vector & output is a vector & embed text for retrieval \\
    \midrule[\heavyrulewidth]
    \multicolumn{5}{@{}l}{\textbf{Consumption edge}: what the program does with the output} \\
    \midrule
    \emph{Group} & \emph{Category} & \emph{Definition} & \emph{Boundary rule} & \emph{Example} \\
    \cmidrule{1-5}
    Control & \texttt{control-flow} & Decides which operations run & picks the path (vs.\ machine-value) & branch or dispatch a tool \\
            & \texttt{executed} & Enters an interpreter & is itself run (vs.\ control-flow) & run generated code or SQL \\
    \cmidrule{1-5}
    Data & \texttt{machine-value} & Parsed into program data & used as data, path unchanged & store a field in state \\
    \cmidrule{1-5}
    Model & \texttt{model-context} & Fed into a prompt or memory & read by a model, not a person & feed the next call \\
          & \texttt{retrieval-index} & Enters a vector index & matched, not read & insert into a vector store \\
    \cmidrule{1-5}
    Human & \texttt{human-display} & Shown to the end user & read by a person & render the answer \\
          & \texttt{evaluation-metric} & Aggregated into a metric & never fed back to control & compute a benchmark score \\
    \bottomrule
  \end{tabular*}
\end{table*}

%% file: tables/screening.tex
\begin{table}[t]
  \caption{Exclusion rules and threat-chain conditions of feasibility screening.}
  \label{tab:screening}
  \centering
  \scriptsize
  \setlength{\tabcolsep}{3pt}
  \renewcommand{\arraystretch}{1.0}
  \begin{tabularx}{\linewidth}{@{}l l X@{}}
    \toprule
    ID & Name & Applies when \\
    \midrule
    \multicolumn{3}{@{}l}{\emph{Exclusion rules}: a gap is excluded if one rule applies} \\
    \cmidrule{1-3}
    E1 & Before the reply & the gap arises before a model reply exists, in the pre-condition or prompt phase \\
    E2 & Shared layer & the defect lies in an extraction, parsing, or retry layer that all outputs share \\
    E3 & Narrowed value & a default-on closed value set, validator, or sanitizer narrows the value to a safe range \\
    E4 & Downstream check & an independent check downstream stops the bad value before it has an effect \\
    E5 & Not live & reaching the defect needs a non-default configuration, or the code is dead \\
    E6 & Crash or quality & the only effect is a failed run or lower quality, with no unauthorized action, injected content, or loss of service \\
    E7 & Unused value & no consumer reads the value, or the value never reaches security-relevant logic \\
    \midrule
    \multicolumn{3}{@{}l}{\emph{Threat chain}: a gap is technically exploitable if all four conditions hold} \\
    \cmidrule{1-3}
    C1 & Defect exists & the code at the analyzed commit has the defect \\
    C2 & Reachable & a real entry point reaches the defect \\
    C3 & Bad value possible & the value can come from attacker input or from a model error \\
    C4 & Consumer acts & the consumer acts on the bad value without further checks \\
    \bottomrule
  \end{tabularx}
\end{table}

%% file: tables/protocol.tex
\begin{table*}[t]
  \caption{Analysis and review protocol with inter-rater agreement.}
  \label{tab:protocol}
  \centering
  \scriptsize
  \setlength{\tabcolsep}{3pt}
  \renewcommand{\arraystretch}{1.0}
  \begin{tabularx}{\textwidth}{@{}c@{\hspace{5pt}}l c X l l r@{}}
    \toprule
     & Step & By & Criterion & Review & Coverage & AC1 ($\kappa$) \\
    \midrule
    \multicolumn{7}{@{}l}{\emph{Calibration}: R1, R2, and R3 refine the guidelines on six pilot systems, then freeze them before the main review} \\
    \midrule
     & Call-site identification & \faCogs\,\faRobot & real LLM invocation, not a wrapper or relay & R1\,$\parallel$\,R2 & all (1,226) & 0.99 (0.98) \\
     & Output decomposition & \faRobot & output split & R1\,$\parallel$\,R2 & all (484) & 0.94 (0.44) \\
     &  &  & operation label & R1\,$\parallel$\,R2 & all (870) & 0.98 (0.98) \\
     & Consumer tracing & \faRobot & consumer location & R1\,$\parallel$\,R2 & sample (885) & 0.91 (0.71) \\
    \multirow{-5}{*}{\grouplabel{Stage I}} &  &  & edge type & R1\,$\parallel$\,R2 & sample (885) & 0.95 (0.94) \\
    \cmidrule{1-7}
     & Responsibility spec. & \faRobot & faithful to the code & R1\,$\parallel$\,R2 & all (537) & 0.94 (0.45) \\
     &  &  & implementation-neutral & R1\,$\parallel$\,R2, code-blind & all (537) & 0.92 (0.33) \\
     & Expected contract & \faRobot & necessary, not too strict, right phase and assurance & R1\,$\parallel$\,R2, code-blind & all (2,438) & 0.95 (0.60) \\
     & Implemented contract & \faRobot & mechanism, phase, assurance match the code & R1\,$\parallel$\,R2 & sample (1,474) & 0.97 (0.57) \\
     & Contract comparison & \faRobot & verdict & R1\,$\parallel$\,R2 & all (4,046) & 0.96 (0.96) \\
    \multirow{-6}{*}{\grouplabel{Stage II}} &  &  & counterexample is valid & R1\,$\parallel$\,R2 & all (4,046) & 0.97 (0.94) \\
    \cmidrule{1-7}
     & Feasibility screening & \faUser & no exclusion rule applies; C1 to C4 hold & -- & all (1,697) & -- \\
     & Gap attribution & \faUser & fixing it breaks or narrows the chain & -- & 290 gaps & -- \\
     & Guided PoC & \faUser & bad value passes default defenses & -- & 60 mechanisms & -- \\
     & Severity grading & \faUser & CVSS vector from the chain evidence & -- & 60 mechanisms & -- \\
     & Trigger testing & \faUser & oracle fixed in advance, repeated runs & fixed oracle & 58 mechanisms & -- \\
    \multirow{-6}{*}{\grouplabel{Stage III}} & Contract-guided fix & \faUser & both payloads blocked, tests pass & -- & 38 mechanisms & -- \\
    \midrule
    \multicolumn{7}{@{}l}{\emph{Adjudication}: R3 resolves every R1--R2 disagreement (604 for Stages~I and~II).} \\
    \bottomrule
  \end{tabularx}
\end{table*}

%% file: listings/storm-check.tex
\begin{wrapfigure}{r}{0.52\textwidth}
  \vspace{-2pt}
\begin{lstlisting}[style=paper, firstnumber=733, frame=tb, moredelim={[is][\color{lststr}]{|}{|}}]
|If the topic is appropriate for the engine to process, output |(*@\lstmark{\textcolor{lststr}{"Yes."}}@*)|; otherwise, output "No. The input violates reason [1/2/3/4]".|
|User input: {user_input}"""|(*@\hfill\callout{1}\setcounter{lstnumber}{750}@*)
  response = my_openai_model(prompt)[0](*@\hfill\callout{2}@*)
  if (*@\lstmark{response.\textcolor{lstkw}{startswith}(\textcolor{lststr}{"No"})}@*):(*@\hfill\callout{3}@*)
    ...  # reject: map reason 1-4 to a message(*@\setcounter{lstnumber}{763}@*)
except Exception as e:
  return "Sorry, the input is inappropriate. ..."
return (*@\lstmark{\textcolor{lststr}{"Approved"}}@*)(*@\hfill\callout{4}@*)
\end{lstlisting}
  \vspace{-2pt}
  \caption{A \texttt{classify} gate in STORM (abridged).}
  \label{fig:storm-check}
  \vspace{-\intextsep}
\end{wrapfigure}

%% file: tables/roles.tex
\begin{table*}[t]
  \caption{Recurring roles of LLM call sites.}
  \label{tab:roles}
  \centering
  \scriptsize
  \setlength{\tabcolsep}{3pt}
  \renewcommand{\arraystretch}{1.0}
  \begin{tabularx}{\textwidth}{@{}l l r r r >{\raggedright\arraybackslash}X@{}}
    \toprule
    Role & Operation $\rightarrow$ consumer & Out. & Sites & Sys. & Example \\
    \midrule
    \multicolumn{6}{@{}l}{\emph{Steering the program}: the output decides which operations run, or is itself run} \\
    \cmidrule{1-6}
    Action selection & \texttt{select} $\rightarrow$ \texttt{control-flow} & 140 & 117 & 26 & Codex: the chosen tool call is dispatched and run \\
    Gating and judging & \texttt{classify}/\texttt{score} $\rightarrow$ \texttt{control-flow} & 72 & 50 & 17 & Gemini CLI: a policy verdict blocks a tool call \\
    Free-text steering & \texttt{generate} $\rightarrow$ \texttt{control-flow} & 139 & 126 & 20 & Aider: files named in the reply join the chat \\
    Code execution & \texttt{generate} $\rightarrow$ \texttt{executed} & 66 & 66 & 9 & Khoj: generated Python code runs in a sandbox \\
    \midrule
    \multicolumn{6}{@{}l}{\emph{Feeding other components}: the output becomes data, context, or a message to a person} \\
    \cmidrule{1-6}
    Model-to-model hand-off & \texttt{generate} $\rightarrow$ \texttt{model-context} & 227 & 191 & 28 & TradingAgents: bull and bear rebut each other \\
    Context compression & \texttt{compress} $\rightarrow$ \texttt{model-context} & 46 & 46 & 15 & Cline: a summary replaces the chat history \\
    Generated program data & \texttt{generate} $\rightarrow$ \texttt{machine-value} & 320 & 278 & 38 & TradingAgents: prices become trade fields \\
    Structured extraction & \texttt{extract} $\rightarrow$ \texttt{machine-value} & 33 & 33 & 9 & DeerFlow: the reply becomes memory updates \\
    User-facing answer & \texttt{generate} $\rightarrow$ \texttt{human-display} & 162 & 142 & 29 & OpenCode: the reply is shown in the terminal \\
    Delegation & \texttt{delegate} $\rightarrow$ any consumer & 19 & 19 & 8 & Gemini CLI: a task goes to a remote agent \\
    \bottomrule
  \end{tabularx}
\end{table*}

%% file: tables/mechanisms.tex
\begin{table*}[t]
  \caption{Code-level mechanisms behind implemented constraints.}
  \label{tab:mechanisms}
  \centering
  \scriptsize
  \setlength{\tabcolsep}{3pt}
  \renewcommand{\arraystretch}{1.0}
  \begin{tabularx}{\textwidth}{@{}l l r r >{\raggedright\arraybackslash}X@{}}
    \toprule
    Mechanism & What it enforces & Constr. & Sys. & Example \\
    \midrule
    \multicolumn{5}{@{}l}{\emph{Before the call}: what the model receives and what it may do} \\
    \cmidrule{1-5}
    Capability restriction & which tools and models are usable & 207 & 32 & TradingAgents: analyst bound to four tools \\
    Configuration checks & call only in an enabled setup & 355 & 34 & Gemini CLI: router runs only if enabled \\
    State checks & call only on meaningful input & 344 & 34 & Cline: no request for an empty diff \\
    Input size limits & input within length bounds & 122 & 25 & OpenClaw: tool-call IDs cut to 64 chars \\
    Input formatting & input in the expected shape & 341 & 32 & UFO: one system and one user message \\
    \midrule
    \multicolumn{5}{@{}l}{\emph{In the request}: how the model is asked} \\
    \cmidrule{1-5}
    Structured output request & reply follows a schema & 59 & 19 & TradingAgents: typed sentiment report \\
    Fixed request parameters & fixed sampling and limits & 294 & 35 & STORM: temperature 0 by default \\
    Request assembly & untrusted text delimited & 179 & 29 & DeerFlow: messages escaped, wrapped in tags \\
    \midrule
    \multicolumn{5}{@{}l}{\emph{After the call}: what the program accepts} \\
    \cmidrule{1-5}
    Schema and type validation & reply parses to a type & 284 & 32 & UFO: reply validated by a Pydantic model \\
    Allow-list and range checks & values in a set or range & 114 & 24 & OpenHands: risk score cut into three levels \\
    Extraction and normalization & usable part isolated & 557 & 36 & DeerFlow: think blocks and fences stripped \\
    Stream handling & stream assembled correctly & 228 & 27 & Cline: only text chunks forwarded \\
    Error handling and retry & failures retried or raised & 479 & 38 & UFO: retries with exponential backoff \\
    Downstream gating & output approved before use & 185 & 27 & Cline: tool calls need user approval \\
    \bottomrule
  \end{tabularx}
\end{table*}

%% file: tables/gap-patterns.tex
\begin{table*}[t]
  \caption{Recurring gap patterns between expected and implemented contracts.}
  \label{tab:gap-patterns}
  \centering
  \scriptsize
  \setlength{\tabcolsep}{3pt}
  \renewcommand{\arraystretch}{1.0}
  \begin{tabularx}{\textwidth}{@{}l l@{\hspace{2pt}$\rightarrow$\hspace{2pt}}l r r r >{\raggedright\arraybackslash}X@{}}
    \toprule
    Pattern & Expected & Implemented & Gaps & Sites & Sys. & Example \\
    \midrule
    \multicolumn{7}{@{}p{\textwidth}@{}}{\emph{Before the call}: what the model receives and what it may do \hfill 420 gaps (25\%)} \\
    \cmidrule{1-7}
    Unchecked call inputs & inputs checked & call made anyway & 330 & 265 & 34 & Continue: prompt not checked before call \\
    Over-broad authority & least authority & wider tool set & 90 & 74 & 27 & OpenManus: planner may also delete plans \\
    \midrule
    \multicolumn{7}{@{}p{\textwidth}@{}}{\emph{In the request}: how the model is asked \hfill 335 gaps (20\%)} \\
    \cmidrule{1-7}
    Underspecified prompt & task fully stated & left unstated & 242 & 168 & 32 & CodeWhale: router not told request is data \\
    Faulty request assembly & input intact, fenced & altered or unfenced & 93 & 88 & 27 & PentAGI: raw tool output inlined unescaped \\
    \midrule
    \multicolumn{7}{@{}p{\textwidth}@{}}{\emph{After the call}: what the program accepts \hfill 829 gaps (49\%)} \\
    \cmidrule{1-7}
    Partial format check & well-formed reply & partly checked & 249 & 218 & 32 & STORM: empty section written to article \\
    Unchecked content & grounded reply & accepted as is & 253 & 199 & 34 & agenticSeek: summary kept even if empty \\
    Fail-open handling & invalid reply rejected & default taken & 146 & 128 & 30 & JARVIS: no ``yes'' means no tool call \\
    Unvalidated action & action in offered set & run as proposed & 138 & 103 & 28 & Goose: unoffered tool names accepted \\
    Prompt-only check & checked in code & only prompted & 43 & 33 & 12 & Gemini CLI: ``no change'' flag trusted \\
    \midrule
    \multicolumn{7}{@{}p{\textwidth}@{}}{\emph{Across paths}: whether a check holds everywhere \hfill 101 gaps (6\%)} \\
    \cmidrule{1-7}
    Path-dependent check & checked on all paths & main path only & 101 & 81 & 28 & SWE-agent: one-action check on one parser \\
    \bottomrule
  \end{tabularx}
\end{table*}

%% file: listings/conseca-failopen.tex
\begin{lstlisting}[style=paper, language=Java, morekeywords={const,await}, firstnumber=36, frame=tb, moredelim={[is][\color{lststr}]{|}{|}}]
|   - "decision": "allow", "deny", or "ask_user".|(*@\hfill\callout{1}\setcounter{lstnumber}{78}@*)
... = JSON.stringify((*@\lstmark{policy[toolName] || \{\}}@*), ...);(*@\hfill\callout{2}\setcounter{lstnumber}{87}@*)
try {
  const result = await ...generateContent(...);(*@\setcounter{lstnumber}{118}@*)
  if (!responseText)
    return { decision: (*@\lstmark{SafetyCheckDecision.ALLOW}@*), ...};(*@\hfill\callout{3}\setcounter{lstnumber}{126}@*)
  try {
    const parsed = EnforcementResultSchema.parse(...);(*@\setcounter{lstnumber}{131}@*)
    switch (parsed.decision) {(*@\setcounter{lstnumber}{138}@*)
      case 'deny': default:(*@\setcounter{lstnumber}{140}@*)
        decision = (*@\lstmark{SafetyCheckDecision.DENY}@*);(*@\setcounter{lstnumber}{148}@*)
  } catch (parseError) {
    return { decision: (*@\lstmark{SafetyCheckDecision.ALLOW}@*), ...};(*@\hfill\callout{4}\setcounter{lstnumber}{155}@*)
} catch (error) {(*@\setcounter{lstnumber}{158}@*)
  return { decision: (*@\lstmark{SafetyCheckDecision.ALLOW}@*), ...};(*@\hfill\callout{5}@*)
\end{lstlisting}
  \vspace{2pt}
  \caption{A fail-open safety check in Gemini CLI (abridged).}
  \label{fig:conseca}

%% file: figures/case-gptresearcher.tex
\begingroup
\definecolor{advred}{HTML}{B3261E}
\definecolor{gapmiss}{HTML}{B3261E}
\definecolor{gappart}{HTML}{A8620A}
\definecolor{sysblue}{HTML}{1B4F72}
\definecolor{sysbg}{HTML}{EFF4F9}
\definecolor{mutegray}{HTML}{6E6E6E}
\newcommand{\cdB}[1]{{\ttfamily\fontsize{6.2}{7.2}\selectfont #1}}
\newcommand{\reqlblB}[1]{{\fontsize{5.6}{5.6}\selectfont\bfseries\color{mutegray}#1}}
\newcommand{\chipB}[2]{%
  \tikz[baseline=(c.base)]{\node[inner xsep=1.6pt, inner ysep=0.5pt,
    rounded corners=0.8pt, fill=#1!10, draw=#1, line width=0.3pt, text=#1,
    font=\fontsize{5.6}{5.6}\selectfont\bfseries] (c) {#2};}}
\newcommand{\passesB}[1]{%
  \\[0.7pt]{\color{sysblue!40}\rule{\linewidth}{0.25pt}}\\[-3.3pt]
  {\color{sysblue}\fontsize{5.6}{5.6}\selectfont\bfseries PASSES ON}~#1}
\begin{tikzpicture}[
  font=\fontsize{6.8}{8.0}\selectfont,
  stageB/.style={draw=sysblue, fill=sysbg, rounded corners=1.6pt, line width=0.45pt,
                anchor=north, inner xsep=3.5pt, inner ysep=2.6pt,
                text width=54mm, align=left},
  srcB/.style={stageB, draw=advred, fill=advred!6},
  hitB/.style={stageB, draw=advred, fill=advred!6, line width=0.8pt},
  flowB/.style={-{Stealth[length=3.4pt,width=2.8pt]}, line width=0.8pt, draw=mutegray},
]
\def\bx{0mm}
\def\fx{-23mm}

\node[srcB] (gsrc) at (\bx,0) {%
  \textcolor{advred}{\bfseries Untrusted input}\hfill\cdB{data.content}\\[0.5pt]
  a chat reply delivered over the websocket, LLM- or web-derived text%
};

\node[stageB, below=3.2mm of gsrc.south, anchor=north] (g1) {%
  \textbf{1\,\,Choose a render path}%
  \hfill{\color{mutegray}\cdB{scripts.js}}\\[0.7pt]
  \reqlblB{REQ}~every render of model output is sanitized first\\[0.7pt]
  \chipB{gapmiss}{MISSING}~\cdB{addChatMessage()} skips \cdB{sanitizeHtml()}, unlike its two siblings
  \passesB{\cdB{showdown} HTML with an unescaped \cdB{onerror} attribute}%
};

\node[hitB, below=3.2mm of g1.south, anchor=north] (g2) {%
  \textbf{2\,\,Insert into the page}%
  \hfill{\color{mutegray}\cdB{innerHTML}}\\[0.7pt]
  \reqlblB{REQ}~\textcolor{red}{scripts in rendered content must not run}\\[0.7pt]
  \chipB{gapmiss}{MISSING}~no sanitizer, no CSP at this call site
  \passesB{\textcolor{advred}{\bfseries observed:} \cdB{onerror} fires in a real browser at the app's own origin}%
};

\foreach \a/\b in {gsrc/g1, g1/g2}
  \draw[flowB] (\fx,0 |- \a.south) -- (\fx,0 |- \b.north);

\end{tikzpicture}
\endgroup

%% file: figures/case-khoj.tex
\begingroup
\definecolor{advred}{HTML}{B3261E}
\definecolor{gapmiss}{HTML}{B3261E}
\definecolor{gappart}{HTML}{A8620A}
\definecolor{sysblue}{HTML}{1B4F72}
\definecolor{sysbg}{HTML}{EFF4F9}
\definecolor{mutegray}{HTML}{6E6E6E}
\newcommand{\cdC}[1]{{\ttfamily\fontsize{6.2}{7.2}\selectfont #1}}
\newcommand{\reqlblC}[1]{{\fontsize{5.6}{5.6}\selectfont\bfseries\color{mutegray}#1}}
\newcommand{\chipC}[2]{%
  \tikz[baseline=(c.base)]{\node[inner xsep=1.6pt, inner ysep=0.5pt,
    rounded corners=0.8pt, fill=#1!10, draw=#1, line width=0.3pt, text=#1,
    font=\fontsize{5.6}{5.6}\selectfont\bfseries] (c) {#2};}}
\newcommand{\passesC}[1]{%
  \\[0.7pt]{\color{sysblue!40}\rule{\linewidth}{0.25pt}}\\[-3.3pt]
  {\color{sysblue}\fontsize{5.6}{5.6}\selectfont\bfseries PASSES ON}~#1}
\begin{tikzpicture}[
  font=\fontsize{6.8}{8.0}\selectfont,
  stageC/.style={draw=sysblue, fill=sysbg, rounded corners=1.6pt, line width=0.45pt,
                anchor=north, inner xsep=3.5pt, inner ysep=2.6pt,
                text width=54mm, align=left},
  hitC/.style={stageC, draw=advred, fill=advred!6, line width=0.8pt},
  flowC/.style={-{Stealth[length=3.4pt,width=2.8pt]}, line width=0.8pt, draw=mutegray},
]
\def\bx{0mm}
\def\fx{-23mm}

\node[stageC] (ksrc) at (\bx,0) {%
  \textbf{Ordinary input}\hfill\cdC{``summarize http://127.0.0.1:...''}\\[0.5pt]
  any authenticated user's own chat message, naming a URL; no injection needed%
};

\node[stageC, below=3.2mm of ksrc.south, anchor=north] (k1) {%
  \textbf{1\,\,Select a scraper for the URL}%
  \hfill{\color{mutegray}\cdC{online\_search.py}}\\[0.7pt]
  \reqlblC{REQ}~internal URLs are rejected or admin-only\\[0.7pt]
  \chipC{gapmiss}{MISSING}~\cdC{is\_internal\_url()} \textbf{routes} internal URLs to \cdC{Direct}, which serves every user
  \passesC{the internal URL, routed to the \cdC{Direct} scraper}%
};

\node[hitC, below=3.2mm of k1.south, anchor=north] (k2) {%
  \textbf{2\,\,Fetch the URL}%
  \hfill{\color{mutegray}\cdC{read\_webpage\_at\_url()}}\\[0.7pt]
  \reqlblC{REQ}~\textcolor{red}{fetches are restricted to an allowlist}\\[0.7pt]
  \chipC{gapmiss}{MISSING}~plain \cdC{aiohttp} GET with no allowlist
  \passesC{\textcolor{advred}{\bfseries observed:} the loopback canary is fetched and returned in the user's reply}%
};

\foreach \a/\b in {ksrc/k1, k1/k2}
  \draw[flowC] (\fx,0 |- \a.south) -- (\fx,0 |- \b.north);

\end{tikzpicture}
\endgroup